\documentclass{article}
\usepackage{graphicx}
\usepackage{amsmath}

\begin{document}

\title{Transport and dynamics on open quantum graphs}
\author{F. Barra\footnote{present address: Chemical Physics department.
Weizmann Institute of Science. Rehovot 76100 Israel.}~ and P. Gaspard\\{\em Center for Nonlinear Phenomena and Complex
Systems,} \\{\em Universit\'{e} Libre de Bruxelles, Campus Plaine C.P. 231,}\\
{\em B-1050 Brussels, Belgium.}}

\maketitle

\begin{abstract}
We study the classical limit of quantum mechanics on graphs
by introducing a Wigner function for graphs.
The classical dynamics is compared to
the quantum dynamics obtained from
the propagator. In particular we consider extended open
graphs whose classical dynamics generate a  diffusion process.
The transport properties of the classical system are revealed
in the scattering resonances and in the time evolution
of the quantum system. 
\end{abstract}

\section{Introduction}

In this article we study quantum properties of systems which present 
transport behavior such as normal diffusion in the classical limit. 
For classical systems, transport phenomena has been related to 
dynamical quantities by the escape rate formalism \cite{Gasp98}. In this formalism,
the escape rate given by the leading Pollicott-Ruelle
resonance, determines the diffusion coefficient of the system in the
large-system limit. Since the classical time evolution is a good
approximation of the quantum evolution in the
so-called semiclassical regime, one expects that kinetic properties such as
the escape rate and the diffusion coefficient will emerge out of the
quantum dynamics.  Connections between the
quantum scattering resonances and the classical diffusive behavior are also
expected because, for open quantum systems, the quantum scattering resonances
determine the time evolution of the wavefunction.

The purpose of the present paper is to explore these kinetic phenomena in
model systems known as quantum  graphs. These systems have similar spectral
statistics of  energy levels as the classically chaotic systems 
\cite{Smilansky0,Smilansky1}. 
Since the pioneering work by Kottos and Smilansky, several studies
have been devoted to the spectral properties of quantum graphs 
\cite{Smilansky2,Keating, art3} and to their applications in mesoscopic physics
\cite{akkermans}. 
We have studied the level
spacing distribution of quantum graphs getting some exact results
in simple cases \cite{art3}.
They have also provided the first model with
a semiclassical description of Anderson localization \cite{Schanz}.
Moreover, the classical dynamics on these systems
has been studied in detail in Ref. \cite{art4} where we have introduced a
time-continuous classical dynamics. In this way, the quantum and classical
dynamics on graphs -- and the relationships between both -- can be studied on
the same ground as in other systems like billiards for instance. In this
article, we go further in the connection between the quantum and classical
dynamics by showing that the classical dynamics of Ref. \cite{art4} emerges
out of the quantum dynamics introduced in Refs.
\cite{Smilansky0,Smilansky1}.

With this correspondence established, and thanks to their simplicity,
the quantum graphs turn out to be good models to study the quantum properties
of systems which present transport properties such as diffusion in the
classical limit.  These properties have been previously studied in
systems like the kicked rotor which has a classical dynamics (given by the
standard map) presenting deterministic diffusion for some values of the
parameters.  In particular, the decay of the quantum staying probability, 
for an open version of the kicked rotor, has been compared to the classical 
decay obtained numerically from the simulation of trajectories of the 
corresponding open standard map \cite{Casati}. 
It has been argued that the decay observed in the quantum
staying probability is determined by the Pollicott-Ruelle 
resonances but no direct evidence has been reported. 
Here, we present results that show that this is indeed the case.
The continuous time evolution of the wavefunction for graphs is 
obtained from the propagator which is obtained from the Fourier 
transform of the Green function on the graph.
We present here a derivation that allows us to compute this Green function
and, therefore, the propagator. On the other hand the classical
dynamics on graphs developed in Ref. \cite{art4} allows us to compute
the Pollicott-Ruelle resonances.  The Pollicott-Ruelle resonances are of
special importance because their quantum manifestation has been found in
experimental measures of some correlation functions in microwave
scattering \cite{Sridhar}.

Apart from the aforementioned time-dependent quantities we have also
studied spectral quantities like the quantum scattering resonances.
The quantum scattering resonances have also been studied for the open
kicked rotor. The distribution of their imaginary part has been 
conjectured to be related to the diffusion process observed
for the classical system \cite{Guarnieri}. Here, we present numerical support
for this conjecture by showing that the widths of the quantum resonances have
the power-law distribution of Ref. \cite{Guarnieri} for some multiconnected
diffusive graphs.

The article is organized as follows. In Section \ref{sec.quan.dyn},
we define the quantum graphs and we review some of their main already known
properties, namely, the formulation of their quantization
and its exact trace formula. Section \ref{quant.scatt.sec} presents the
problem of scattering on quantum graphs. 
In Subsection \ref{green.sec}, we introduce a multi-scattering expansion
for the Green function on graphs. Green function on graphs
have been considered elsewhere \cite{akkermans}, but to our knowledge
the multi-scattering expression and the resumed closed form that we obtain
are new results. The knowledge of the Green function allows us to
obtain the propagator on graphs and therefore to have access to time-dependent
phenomena on graphs. The propagator is introduced in Subsection
\ref{propa.sec}. The emergence of the classical dynamics out of the quantum
dynamics is studied in Section \ref{classical.sec}.  In Subsection
\ref{wigner.graph.sec}, we introduce a  Wigner function for graphs and we
compute the classical limit by neglecting interference between different
paths. This limit corresponds to the classical probability density on the
graphs as we show in Subsection
\ref{sec.class.dyn} where we also summarize the most 
important results about the classical theory of graphs \cite{art4}. 
The quantum time evolution
of open graphs is considered in Subsection \ref{quant.evol.sec} in which
we compare the decay of the quantum staying probability with
the classical decay of the density as obtained from the Pollicott-Ruelle
resonances presented in Subsection \ref{sec.P.R.res}.  We do this comparison
for small systems and for large systems  where a diffusion process dominate the
classical escape. In Section \ref{sec.quantum.res}, we analyze the statistical
properties of the distribution of quantum scattering resonances. Here, we
present a new derivation of the resonance density using the concept of
Lagrange mean motion of an almost-periodic function.  Examples are given and
discussed in Section \ref{sec.ejem}.  Conclusions are drawn in Section
\ref{conclu.sec}.

\section{Quantum graphs}

\label{sec.quan.dyn}

\subsection{Definition of graphs}
\label{sec.def}

Let us introduce graphs as geometrical objects where a particle moves.
Graphs are $V$ vertices connected by $B$ bonds.
Each bond $b$ connects two vertices,
$i$ and $j$. We can assign an orientation to each bond and define ``oriented
or directed bonds''. Here, one fixes the direction of the bond $[i,j]$ and
call $b=(i,j)$ the bond oriented from $i$ to $j$. The same bond but oriented
from $j$ to $i$ is denoted $\hat{b}=(j,i)$.  We notice that
$\hat{\hat{b}}=b$.  A graph with $B$ bonds has $2B$ directed bonds.
The valence $\nu_i$ of a vertex is the number of bonds that meet at the
vertex $i$.

Metric information is introduced by assigning a length $l_b$ to each bond $b$.
In order to define the position of a particle on the graph, we introduce a
coordinate $x_b$ on each bond $b=[i,j]$. We can assign either the
coordinate $x_{(i,j)}$ or
$x_{(j,i)}$. The first one is defined such that $x_{(i,j)}=0$ at $i$ and
$x_{(i,j)}=l_b$ at
$j$, whereas $x_{(j,i)}=0$ at $j$ and $x_{(j,i)}=l_b$ at $i$. Once the
orientation is
given, the position of a particle on the graph is determined by the
coordinate $x_{b}$ where $0\leq x_{b}\leq l_{b}$. The index $b$ identifies
the bond and the value of $x_{b}$ the position on this bond.

For some purposes, it is convenient to consider $b$ and $\hat{b}$ as
different bonds within the formalism. Of course, the physical quantities
defined on each of them must satisfy some consistency relations. In particular,
we should have that $l_{\hat{b}}=l_{b}$ and $x_{\hat{b}}=l_{b}-x_{b}$.

We introduce here some notations that we are going to use next. For oriented
bonds $b$ we define the functions $q(b)$ and $p(b)$ which give the vertex
at the origin and at the end of $b$, respectively. Thus, for the bond
$b=(i,j)$, we have $q(b)=i$ and $p(b)=j$. These functions are well defined
for graphs with multiple loops and bonds also. In the last case these
functions take the same values for two or more different bonds. Note that we
have the following equalities $p(b)=q(\hat{b})$ and $q(b)=p(\hat{b})$.

\section{Quantum mechanics of a particle on a graph}

On each bond $b$, the component $\psi_{b}$ of the total wavefunction
$\pmb{\Psi}$ is a solution of the one-dimensional Schr\"{o}dinger equation.
This means that the dimension of the vector $\pmb{\Psi} =[\psi _{1}(x),\ldots
,\psi _{B}(x)]^{
\mathrm{T}}$ is $B$, but when we consider directed bonds as different it
will be of dimension $2B$, with $B$ components containing redundant
information [see Eq.(\ref{symmetry}) below]. We consider the time-reversible
case (i.e., without magnetic field)
\begin{equation}
-\frac{d^{2}}{dx^{2}}\psi _{b}(x)=k^{2}\psi _{b}(x),\qquad b=(i,j)\ ,
\label{schr.graph}
\end{equation}
where $k=\sqrt{2mE}/\hbar$ is the wavenumber and $E$ the
energy. [We use the short hand notation $\psi _{b}(x)$ for 
$\psi _{b}(x_{b})$ and is understood that $x$ is the
coordinate on the bond $b$ to which the component $\psi _{b}$ refers.]
Moreover, the wavefunction must satisfy boundary conditions at the
vertices of each bond ($i$ and $j$ in the previous equation). The solutions
will have the form 
\begin{equation}
\psi _{b}(x)=\psi _{+}(b)\exp (ikx)+\psi _{-}(b)\exp (-ikx)
\end{equation}
where the boundary conditions impose restrictions on $\psi
_{+}(b)$ and $\psi _{-}(b)$ which are the amplitudes of the forward and
backward moving waves on the bond $b$. 

If we consider oriented bonds, the wavefunction in a bond $b$ and in the
corresponding reverted bond $\hat{b}$ must satisfy the following
consistency relation 
\begin{equation}
\psi _{\hat{b}}(x)=\psi _{b}(l_{b}-x).  \label{symmetry}
\end{equation}

\subsection{Boundary conditions and quantization conditions}

\subsubsection{Vertex matrix}

A natural boundary condition is to impose the continuity of the wavefunction at
all the vertices together with current conservation, i.e., 
\begin{equation*}
\psi _{b}(0)=\varphi _{i}
\end{equation*}
for all the bonds $b$ which start at the vertex $i$ and 
\begin{equation*}
\psi _{b}(l_{b})=\varphi _{j}
\end{equation*}
for all the bonds $b$ which end in the vertex $j$. The ``current
conservation'' reads 
\begin{equation}
\sum_{b}\delta \left[ q(b),i\right]\frac{d}{dx}\psi
_{b}(x)\Big\vert_{x\rightarrow 0}=-\alpha _{i}\ \varphi _{i} 
\label{current.cons}
\end{equation}
where $\sum_{b}$ denotes a summation over all the $2B$ directed bonds.
The Kronecker delta function selects those bonds which have their origin at
the vertex $i$ and we impose this condition at all vertex $i$. 
The case where $\alpha_{i}=0$ is referred to as Neumann boundary condition. 
For nonvanishing $\alpha_{i}$, Eq.(\ref {current.cons}) is the most general
boundary condition for which the resulting Schr\"{o}dinger operator is
self-adjoint \cite{seba1,exner}. In the case where the vertex $i$ connects two
bonds, $\alpha_{i}$ plays the role of the intensity of a delta potential
localized at the position of the vertex. In fact, Eq.(\ref{current.cons})
represents the discontinuity of the derivative of the wavefunction at the
point where the delta potential is located. Due to this analogy we call the
$\alpha _{i}$ the vertex potentials.

This boundary condition gives \cite{Smilansky1,AvronRMP}
\begin{equation}
\sum_{j\in V}h_{ij}\varphi_j=0  \label{VxV.syst.eq}
\end{equation}
where the sum is over all the vertices of the graph and the vertex-vertex
matrix $\mathsf h$ is defined by 
\begin{equation*}
h_{ij}=-\left\{ \sum_{b}\frac{\delta [q(b),i]\delta [p(b),j]}{\sin kl_{b}}
\right\} +\left\{ \sum_{b}\delta [q(b),i]\; \cot kl_{b}-\frac{\alpha _{i}}{k}
\right\} \delta (i,j)
\end{equation*}
This expression for $\mathsf h$ is valid for graphs with loops and multiple
bonds connecting two vertices. Note that a loop contributes twice to each sum.
In the absence of loops and multiple bonds it simplifies to 
\begin{equation}
h_{ij}=\left\{ 
\begin{array}{cc}
\delta (i,j)\sum_{b}\left\{ \delta [q(b),i]\; \cot
kl_{b}-\frac{\alpha_{i}}{k}
\right\}\; , & i=j \\  &  \\ 
-C_{ij}\; \frac{1}{\sin kl_{b}}\; , & i\neq j
\end{array}
\right.  \label{M h}
\end{equation}
where $C_{ij}$ is called the connectivity matrix.
Eq.(\ref{VxV.syst.eq}) has solutions only if 
\begin{equation*}
\det {\mathsf h}(k)=0\; .  
\end{equation*}
This is the secular equation that gives the eigenenergies $\left\{
k_{n}^{2}\right\} $ of the graph. This equation can be solved numerically to
obtain an arbitrary number of eigenenergies. Therefore, graphs are very nice
systems to study the statistics of the spectrum fluctuations.

\subsubsection{Bond matrix}

\label{sec.BondMatrix}

There is however a general boundary condition, where we consider a
scattering process at each vertex. In each vertex a (unitary) scattering
matrix relates outgoing waves to the incoming ones. 
If we call $\pmb{\sigma}^{i}$ the
scattering matrix for the vertex $i$ the condition is 
\begin{equation*}
\psi _{a}^{\rm out}(i)=\sum_{b}\sigma_{ab}^{i}\psi _{b}^{\rm in}(i)
\end{equation*}
where the sum is over all the (non-directed) bonds that meet at $i$. 
For $a\neq b$, $\sigma
_{ab}^{i}$ is the transmission amplitude for a wave that is incident at the
vertex $i$ from the bond $b$ and is transmitted to the bond $a$. Similarly, 
$\sigma _{aa}^{i}$ is a reflection amplitude. The matrix $\pmb{\sigma}^{i}$ has
dimension $\nu_{i}\times \nu_{i}$ where $\nu_{i}$ is the valence of 
the vertex $i$. This equation is imposed at all vertices and for all the bonds
that meet in the vertex.

Now we consider oriented bonds which allows us to
write the previous boundary condition in the following way 
\begin{equation}
\psi _{a}^{\rm out}[q(a)]=\sum_{b}\sigma _{ab}^{q(a)}\delta [q(a),p(b)]\psi
_{b}^{\rm in}[p(b)]  \label{bc.directed}
\end{equation}
where the sum is over all the $2B$ directed bonds and the equation is
imposed on every directed bond $a$ because the set of all directed bonds is
equivalent to the set of all vertices with the (non-directed) bonds that
meet at the vertex.

As a consequence of Eq.(\ref{symmetry})
we have the relation 
\begin{equation}
\psi _{b}^{\rm in}[p(b)]=\exp
(ikl_{b})\psi _{b}^{\rm out}[q(b)]\; .
\label{prop.bond}
\end{equation} 
Setting the expression (\ref{prop.bond}) in Eq.(\ref
{bc.directed}) we get 
\begin{equation*}
\psi _{a}^{\rm out}[q(a)]=\sum_{b}T_{ab}\exp (ikl_{b})\psi _{b}^{\rm out}[q(b)]
\end{equation*}
from which follows the quantization condition 
\begin{equation}
\det [{\mathsf I}-{\mathsf R}(k)]=0  \label{quant1}
\end{equation}
with 
\begin{equation}
{\mathsf R}={\mathsf T}{\mathsf D}(k)  \label{quant1a}
\end{equation}
a unitary matrix of dimension $2B$ where 
\begin{equation}
D_{ab}=\delta _{ab}\ e^{ikl_{a}}\ ,\qquad \text{with}\quad l_{a}=l_{b}
\label{quant1b}
\end{equation}
and 
\begin{equation}
T_{ab}=\sigma _{ab}^{q(a)}\delta [q(a),p(b)]  \label{quant1c}
\end{equation}
Eq.(\ref{quant1}) gives the eigenenergies $\left\{ k_{n}^{2}\right\} .$ Note
that $T_{ab}$ is the transmission amplitude from $b$ to $a$ [if they are
oriented such that $p(b)=q(a)$]. The reflection amplitude is now $T_{
\hat{a}a}$ and not $T_{aa}$ that vanishes due to the delta function in
Eq.(\ref{quant1c}).

The general boundary condition (\ref{bc.directed}) reduces to the previous
one (\ref{current.cons}) for the choice 
\begin{equation}
\sigma _{ab}^{l}=\left( \frac{1+e^{-i\omega _{i}}}{\nu_{l}}-\delta
_{ab}\right) \delta [p(b),l]\delta [q(a),l]
\label{sigma tipic}
\end{equation}
with \cite{Smilansky1} 
\begin{equation}
\omega _{l}=2\arctan\left(\frac{\alpha _{l}}{k\nu_{l}}\right).
\end{equation}

From Eq.(\ref{quant1}) an exact trace formula can be obtained 
\cite{Smilansky0,Smilansky1}
\begin{equation}
d(k)=\frac{L_{\rm tot}}{\pi}+\frac{1}{\pi}\sum_{p,r}
A_p^r l_p \cos(rkl_p)
\label{traceformula}
\end{equation}
where $L_{\rm tot} / \pi$ gives the mean density of levels and the oscillating
term is a sum over prime periodic orbits and their repetitions. 
$A_p=T_{ab}\cdots T_{za}$ is the probability amplitude of the 
prime periodic orbit and
plays the role of stability factor
including the Maslov index. The Lyapunov coefficient per unit 
length of the orbit $\tilde\lambda_p$ is defined by the relation 
$|A_p|^2=e^{\tilde\lambda_p l_p}$.


\section{Quantum scattering on graphs}
\label{quant.scatt.sec}

So far we have considered bonds of finite lengths. When we attach
semi-infinite leads to some vertices, the physical problem changes because
there is now escape from the graph and, thus, it must be analyzed as a
scattering system.  The scattering
matrix is a square matrix of a dimension that equals the number of open
channels. For graphs these channels have a concrete meaning, they are the $L$
semi-infinite leads of the graph. In this section, we shall introduce the
scattering matrix ${\mathsf S}$ for graphs and we shall show that it has a
multi-scattering expansion, closely related to the one we shall obtain for
the Green function. Scattering on graphs was also studied by Kottos and
Smilansky who showed that they display typical features of chaotic
scattering \cite{Smilansky3}.

\subsection{Scattering matrix $\mathsf S$}

On each bond and lead the Schr\"{o}dinger equation allows counter
propagating solutions. We denote by $c$ the scattering leads and by ${\bf
c}=\{c \}$ the set of all the scattering leads.  Moreover, we denote by ${\bf
b}=\{ b\}$ the set of all the bond forming the finite part of the open graph
without its leads.  The
$L\times L$ scattering matrix $\mathsf S$ relates the incoming amplitudes to
the outgoing ones as 
\begin{equation}
\pmb{\Psi}_{\rm out}({\bf c})={\mathsf S}\; \pmb{\Psi}_{\rm in}({\bf c}) 
\label{scatt.relation}
\end{equation}
We shall derive the matrix $\mathsf S$ starting from Eq.(\ref{bc.directed})
which together with Eq.(\ref{quant1c}) reads 
\begin{equation}
\psi _{a}^{\rm out}=\sum_{b}T_{ab}\psi _{b}^{\rm in}  \label{quant1c'}
\end{equation}
This equation is valid for every directed bond and every directed lead. We
have dropped the explicit dependence on the vertex in equation (\ref
{quant1c'}) [see Eq.(\ref{bc.directed})] because it is always understood
that the incoming wave $\psi _{b}^{\rm in}$ in the directed bond $b$ is
incident to the vertex $p(b)$ while the outgoing wave $\psi _{b}^{\rm out}$
emanates from  $q(b)$. This convention is used through all the present paper.
We assume that the scattering leads are oriented from the graph to infinity.
Since the leads are infinite there is no scattering from the lead to the
reverted lead that is the $L\times L$ sub-matrix ${\mathsf T}_{\bf
\hat{c}c}=0$. Neither there is transmission from a bond to a reverted leads
nor from a lead to a bond. That is the $L\times 2B$ sub-matrix
${\mathsf T}_{\bf\hat{c}b}=0$ and the
$2B\times L$ sub-matrix ${\mathsf T}_{\bf bc}=0$.  Moreover,
${\mathsf T}_{\bf\hat{c}\hat{c}}=0$ and
${\mathsf T}_{\bf cc}=0$ due to the delta function in definition
(\ref{quant1c}) and the selection of the orientation of the leads. Thus, in
matrix form, Eq.(\ref{quant1c'}) reads 
\begin{equation*}
\left[ 
\begin{array}{c}
\pmb{\Psi}_{\rm out}(\bf\hat{c}) \\ 
\pmb{\Psi}_{\rm out}(\bf c) \\ 
\pmb{\Psi}_{\rm out}(\bf b)
\end{array}
\right] =\left[ 
\begin{array}{ccc}
0 & 0 & 0 \\ 
{\mathsf T}_{\bf c\hat{c}} & 0 & {\mathsf T}_{\bf cb} \\ 
{\mathsf T}_{\bf b\hat{c}} & 0 & {\mathsf T}_{\bf bb}
\end{array}
\right] \left[ 
\begin{array}{c}
\pmb{\Psi}_{\rm in}(\bf\hat{c}) \\ 
\pmb{\Psi}_{\rm in}(\bf c) \\ 
\pmb{\Psi}_{\rm in}(\bf b)
\end{array}
\right]
\end{equation*}
where ${\mathsf T}_{\bf c\hat{c}}$ is a $L\times L$ matrix whose elements
represent the direct lead-to-lead transmission or
reflection amplitudes. The matrix ${\mathsf T}_{\bf cb}$ is a $L\times 2B$
matrix whose elements represent the bond-to-lead transmission and,
similarly, the matrix ${\mathsf T}_{\bf b\hat{c}}$ is a $2B\times L$ matrix
whose elements represent the lead-to-bond transmission. Finally, the matrix
${\mathsf T}_{\bf bb}$ is the $2B\times 2B$ matrix that we have simply called
${\mathsf T}$ for the closed graphs and represents bond-to-bond transmission or
reflection. Note however that, for a vertex with an attached scattering lead,
the bond-to-bond probability amplitude is different from the one for the graph
without the scattering lead because the valence of the vertex has changed from
$\nu_{i} $ to $\nu_{i}+\#\{$leads attached to $i\}$. 
The previous matrix equation
can be rewritten as 
\begin{eqnarray}
\pmb{\Psi}_{\rm out}({\bf c}) &=&{\mathsf T}_{\bf c\hat{c}} \pmb{\Psi}_{\rm
in}(\hat{\bf c})+{\mathsf T}_{\bf cb} \pmb{\Psi}_{\rm in}({\bf b})
\label{scatt.1} \\
\pmb{\Psi}_{\rm out}({\bf b}) &=&{\mathsf T}_{\bf b\hat{c}} \pmb{\Psi}_{\rm
in}({\bf \hat{c}})+{\mathsf T}_{\bf bb} \pmb{\Psi}_{\rm in}({\bf b})
\label{scatt.2}
\end{eqnarray}
and now we use the relation given in Eq.(\ref{prop.bond}). With
the convention introduced after Eq.(\ref{quant1c'}), we can write Eq.(\ref
{prop.bond}) in a matrix form 
\begin{equation}
\pmb{\Psi}_{\rm in}(b)={\mathsf D}(k) \pmb{\Psi}_{\rm out}(b) 
\label{prop.bond'}
\end{equation}
with the diagonal ($2B\times 2B$) matrix $\mathsf D$ defined in
Eq.(\ref{quant1b}). Thus from Eq.(\ref{scatt.2}) and Eq.(\ref{prop.bond'}) we
get 
\begin{equation*}
\pmb{\Psi} _{\rm in}({\bf
b})=({\mathsf D}^{-1}-{\mathsf T}_{\bf bb})^{-1}{\mathsf
T}_{\bf b\hat{c}}\pmb{\Psi} _{\rm in}(\hat{\bf c})\; .
\end{equation*}
Replacing this last equation in Eq.(\ref{scatt.1}) we obtain 
\begin{equation*}
\pmb{\Psi} _{\rm out}({\bf c})=\left\{
{\mathsf T}_{\bf c\hat{c}}+{\mathsf
T} _{\bf cb}[{\mathsf I}-{\mathsf R}(k)]^{-1}{\mathsf
D}{\mathsf T}_{\bf b\hat{c}}\right\}\pmb{\Psi} _{\rm in}(\hat{\bf c})
\end{equation*}
where we wrote ${\mathsf R}(k)={\mathsf D}(k){\mathsf T}_{\bf bb}$.  That is
the outgoing waves on the leads are determined by the incoming waves on the
leads. This gives the desired scattering matrix 
\begin{equation}
{\mathsf S}={\mathsf T}_{\bf c\hat{c}}+{\mathsf
T}_{\bf cb}[{\mathsf I}-{\mathsf R}(k)]^{-1}{\mathsf D}{\mathsf
T}_{\bf b\hat{c}}
\label{U.matrix}
\end{equation}
which appears in (\ref{scatt.relation}) once we identify each lead and its
reverse as the same physical lead. The multiple scattering expansion is
obtained from 
\begin{equation*}
[ {\mathsf I}-{\mathsf R}(k)]^{-1}=\sum_{n=0}^{\infty }{\mathsf R}(k)^{n}.
\end{equation*}
In a similar way as was done for the trace formula we can get that 
\begin{equation*}
S_{cc^{\prime }}=[{\mathsf T}_{\bf c\hat{c}}]_{cc^{\prime }}+\sum_{p\in
{\cal P}_{c^{\prime }\rightarrow c}}A_{p}\exp [ikl_{p}(c^{\prime },c)]
\end{equation*}
with ${\cal P}_{c^{\prime }\rightarrow c}$ the set of trajectories which goes
from $c^{\prime }$ to $c$. As usual $A_{p}$ is the amplitude of the path and
$l_{p}(c^{\prime },c)$ the length of the path given by the sum of the
traversed bond lengths.

\subsection{Green function for graphs}
\label{green.sec}

Following Balian and Bloch \cite{BalianBlochMultiScatt} we seek for a
multi-scattering expansion of the Green\ function. The Green function
represents the wavefunction in the presence of a point source. We shall
identify the bond where the point source is with the bond $1$. In this
bond, the point source is located at $x_{1}=x^{\prime }$. Since we shall work
with directed bonds the source also appears at the point
$x_{\hat{1}}=l_{1}-x^{\prime }$ on the bond $\hat{1}$. The Green function
satisfies the following equation 
\begin{equation*}
\frac{d^{2}}{dx^{2}}G_{1}(x,x^{\prime })+k^{2}G_{1}(x,x^{\prime })=\frac{2m}{
\hbar ^{2}}\delta (x-x^{\prime })\; ,\qquad 0<x,x^{\prime }<l_{1}
\end{equation*}
\begin{equation*}
\frac{d^{2}}{dx^{2}}G_{\hat{1}}(x,x^{\prime })+k^{2}G_{\hat{1}
}(x,x^{\prime })=\frac{2m}{\hbar ^{2}}\delta (x-l_{1}+x^{\prime })\; ,\qquad
0<x<l_{\hat{1}}=l_{1}
\end{equation*}
\begin{equation*}
\frac{d^{2}}{dx^{2}}G_{b}(x,x^{\prime })+k^{2}G_{b}(x,x^{\prime })=0\; ,\qquad
\forall b\neq \left\{ 1,\hat{1}\right\}
\end{equation*}
Note that $x^{\prime }$ is fixed to belong to the bond $1.$ The Green
function satisfies the same boundary condition as the wavefunction.

In the spirit of a multi-scattering solution we assume that 
\begin{multline}
G_{b}(x,x^{\prime })=G_{0}(x,x^{\prime })\delta
_{b1}+G_{0}(x,l_{1}-x^{\prime })\delta _{\hat{b}1}+ \\
G_{0}(x,0)\mu
_{q(b)}(x^{\prime })+G_{0}(x,l_{b})\mu _{p(b)}(x^{\prime })
\label{green.func}
\end{multline}
where $G_{0}(x,x^{\prime })=\frac{2m}{\hbar ^{2}}\frac{e
^{ik|x-x^{\prime }|}}{2ik}$ is the free Green function and represents an
outgoing wave from $x^{\prime }$. The justification of this ansatz is the
following. In the bond $1$, the wavefunction consists in the superposition
of the wave emanating from the source at $x^{\prime }$ plus waves that
arrive from the borders of the bond. On the other bonds (i.e. not $1$ or $
\hat{1}$) only these waves are present. Since we are dealing with
directed bonds we call the vertices at the border of $b$, $q(b)$ and $p(b)$
as usual.

We have to impose to the Green function the consistency condition 
\begin{equation}
G_{\hat{b}}(x,x^{\prime })=G_{b}(l_{b}-x,x^{\prime }).
\label{condition.green}
\end{equation}
The observation that 
\begin{eqnarray*}
G_{0}(x,l_{1}-x^{\prime }) &=&G_{0}(l_{1}-x,x^{\prime }) \\
G_{0}(l_{b}-x,0) &=&G_{0}(x,l_{b}) \\
G_{0}(l_{b}-x,l_{b}) &=&G_{0}(x,0)
\end{eqnarray*}
leads us to conclude from Eq.(\ref{condition.green}) that 
\begin{eqnarray*}
\mu _{q(\hat{b})}(x^{\prime }) &=&\mu _{p(b)}(x^{\prime }) \\
\mu _{p(\hat{b})}(x^{\prime }) &=&\mu _{q(b)}(x^{\prime })
\end{eqnarray*}
Now, we impose the boundary condition which is given by Eq.(\ref{quant1c'}).
With this aim we have to identify incoming and outgoing components. The
amplitude of the incoming wavefunction from the bond $b$ is 
\begin{equation*}
\psi _{b}^{\rm in}=G_{0}(l_{b},x^{\prime })\delta
_{b1}+G_{0}(l_{b},l_{1}-x^{\prime })\delta _{b\hat{1}}+G_{0}(l_{b},0)\mu
_{q(b)}(x^{\prime })
\end{equation*}
where the first two terms represent the incoming amplitudes of a wave
emanating from $x^{\prime }$ if we are on the bonds $1$ or $\hat{1}$.
The third term is the incoming amplitude of the wave transmitted to the
origin of $b$. For the outgoing amplitude on the bond $b$ we have 
\begin{equation*}
\psi _{b}^{\rm out}=G_{0}(0,0)\mu _{q(b)}(x^{\prime })
\end{equation*}
Thus the boundary condition Eq.(\ref{quant1c'}) gives 
\begin{equation*}
\frac{m}{ik\hbar ^{2}}\mu _{q(a)}(x^{\prime
})=\sum_{b}T_{ab}[G_{0}(l_{b},x^{\prime })\delta _{b1}+
G_{0}(l_{b},l_{1}-x^{\prime })\delta _{b\hat{1}}+G_{0}(l_{b},0)\mu
_{q(b)}(x^{\prime })]
\end{equation*}
where we used that $G_{0}(0,0)=\frac{m}{ik\hbar ^{2}}$. We define $
g(x,x^{\prime })=\frac{ik\hbar ^{2}}{m}G_{0}(x,x^{\prime })$. If we agree
in denoting by ${\mathsf T_g}(l,0)$ the matrix whose $ab$ elements are $
T_{ab}g(l_{b},0) $ and by ${\mathsf T_g}(l,x^{\prime })$ the matrix whose $ab$
elements are $T_{ab}g(l_{b},x^{\prime })$ we can rewrite the previous system of
equations in the more convenient vectorial form 
\begin{equation*}
\pmb{\mu}_{q}(x^{\prime })={\mathsf T_g}(l,x^{\prime })\cdot {\bf e}_{1}+
{\mathsf T_g}(l,l_{1}-x^{\prime })\cdot {\bf e}_{\hat{1}}+{\mathsf
T_g}(l,0)\cdot\pmb{\mu}_{q}(x^{\prime })
\end{equation*}
where ${\bf e}_{1}$ $({\bf e}_{\hat{1}})$ is the $2B$-dimensional vector of
components $[{\bf e}_{1}]_{b}=\delta _{b1}$ $([{\bf e}_{\hat{1}}]_{b}=\delta
_{b\hat{1}}).$ The solution to these equations can be written as 
\begin{equation}
\pmb{\mu}_{q}(x^{\prime }) =\left[{\mathsf I}-{\mathsf
T_g}(l,0)\right]^{-1}\left[ {\mathsf T_g}(l,x^{\prime })\cdot
{\bf e}_{1}+{\mathsf T_g}(l,l_{1}-x^{\prime })\cdot {\bf e}_{\hat{1}}
\right]\; .  \label{mu.q}
\end{equation}
Replacing in (\ref{green.func}) with $\mu_{p(b)}(x^{\prime })=\mu _{q(
\hat{b})}(x^{\prime })$, we have obtained the Green function for graphs.
We note that ${\mathsf T_g}(l,0)={\mathsf R}(k)$ and thus the Green function
has poles at the resonances. The multiple scattering form follows from the
well-known expansion $({\mathsf I}-{\mathsf R})^{-1}={\mathsf
I}+{\mathsf R}+{\mathsf R}^{2}+{\mathsf R}^{3}+\cdots $. Remembering that
$\mu _{p(a)}(x^{\prime })=\mu _{q(\hat{a})}(x^{\prime })$, we have that the
$a$ component of the Green function is 
\begin{align}
G_{a}(x,x^{\prime })& =G_{0}(x,x^{\prime })\delta _{a1}+G_{0}(x,0)\left[
T_{a1}g(l_{1},x^{\prime })+T_{a\hat{1}}g(0,x^{\prime })\right] 
\label{multiscatt} \\
& +G_{0}(x,l_{a})\left[ T_{\hat{a}1}g(l_{1},x^{\prime })+T_{\hat{a}
\hat{1}}g(0,x^{\prime })\right]  \notag \\
& +G_{0}(x,0)\sum_{b}T_{ab}g(l_{b},0)\left[ T_{b1}g(l_{1},x^{\prime })+T_{b
\hat{1}}g(0,x^{\prime })\right]  \notag \\
& +G_{0}(x,l_{a})\sum_{b}T_{\hat{a}b}g(l_{b},0)\left[ T_{b1}g(l_{1},x^{
\prime })+T_{b\hat{1}}g(0,x^{\prime })\right]   \notag \\
& +G_{0}(x,0)\sum_{b,b^{\prime }}T_{ab}g(l_{b},0)T_{bb^{\prime
}}g(l_{b^{\prime }},0)\left[ T_{b^{\prime }1}g(l_{1},x^{\prime
})+T_{b^{\prime }\hat{1}}g(0,x^{\prime })\right]   \notag \\
& +G_{0}(x,l_{a})\sum_{b,b^{\prime }}T_{\hat{a}b}g(l_{b},0)T_{bb^{\prime
}}g(l_{b^{\prime }},0)\left[ T_{b^{\prime }1}g(l_{1},x^{\prime
})+T_{b^{\prime }\hat{1}}g(0,x^{\prime })\right] +\cdots  \notag
\end{align}
Noticing that $g(x,x^{\prime })=e^{ik|x-x^{\prime }|}$ we can write
the previous result in the more handy notation 
\begin{equation}
G_{a}(x,x^{\prime })=\frac{2m}{\hbar ^{2}}\frac{1}{2ik}\sum_{\{p\}}A_{p}
e^{ikl_{p}(x,x^{\prime })}  \label{Green.sum}
\end{equation}
where $A_{p}$ is the probability amplitude of the path $p$ which connects
the initial point $x^{\prime }$ on the bond $1$ to the final point $x$ on
the bond $a$. If the path $p$ is composed by the $n$ bonds $1b_{2}\cdots
b_{n-1}a$ then 
\begin{equation*}
A_{p}=T_{ab_{n-1}}T_{b_{n-1}b_{n-2}}\cdots T_{b_{2}1}
\end{equation*}
The fact that $G_{0}(x,x^{\prime })$ and $g(x,x^{\prime })$ depend on the
modulus of the differences $|x-x^{\prime }|$ implies that, in (\ref
{multiscatt}), we are always adding lengths. Thus $l_{p}(x,x^{\prime })$ is
the total length of the path $s$ which connects $x^{\prime }$ to $x$.

The expression (\ref{Green.sum}) is like a path-integral representation of
the Green function: We add the probability amplitudes of all the paths connecting 
$x^{\prime }$ to $x$ in order to get the Green function.


\subsection{Propagator for graphs}
\label{propa.sec}

Green functions and propagator are
related by Fourier or Laplace transforms: 
\begin{equation*}
G_{a}^{(+)}(x,x^{\prime };E)=\lim_{\epsilon \rightarrow 0^{+}}\frac{1}{
i\hbar }\int_{0}^{\infty }dt\exp (-\epsilon t)\exp \left(\frac{i}{\hbar }
Et\right)K_{a}(x,x^{\prime };t)
\end{equation*}
\begin{equation}
K_{a}(x,x^{\prime };t)=\frac{1}{2\pi
i}\int_{{\cal C}_{+}+{\cal C}_{-}}dE\exp\left(-\frac{i}{
\hbar }Et\right)G_{a}(x,x^{\prime };E)  \label{propagator}
\end{equation}
where the contours ${\cal C}_{+}$ goes from $\mathrm{Re}\,E=+\infty$
to $\mathrm{Re}\,E=-\infty$ with a positive imaginary part, while ${\cal
C}_{-}$ goes from $\mathrm{Re}\,E=-\infty$
to $\mathrm{Re}\,E=+\infty$ with a negative imaginary part.

Using expression (\ref{Green.sum}) for the Green
function, we get from Eq.(\ref{propagator}) that 
\begin{equation}
K_{a}(x,x^{\prime };t)=\sqrt{\frac{m}{2i\pi \hbar t}}\sum_{\{p\}}A_{p}\;
e^{i\frac{ml_{p}(x,x^{\prime })^{2}}{2\hbar t}}  \label{prop.path}
\end{equation}
This expression shows that the propagator is the sum over all the paths $\{p\}$
that join $x^{\prime }$ to $x$ in a fixed time $t$.
Each term is composed of a free propagator 
weighted by the probability amplitude of the given path. This
result could have been guessed from the general principles of quantum
mechanics, i.e., if there are many ways to obtain a given result then the
probability amplitude is the sum of the probability amplitudes of the
different ways of obtaining the result.

The closed form of the Green function [given
by Eqs.(\ref{green.func}) \& (\ref{mu.q})] and the fast Fourier transform
allow us to obtain numerically the propagator as a function of the time $t$
and, therefore, the time evolution of a wave packet. We shall develop this
possibility after analyzing the classical limit of quantum mechanics on graphs
in the following section.


\section{Emerging classical dynamics on quantum graphs}
\label{classical.sec}

The emergence of the classical dynamics out of the quantum dynamics can be
studied by introducing the concept of Wigner function.  Such a function should
tend to the classical probability density in the classical limit.

\subsection{Wigner functions on graphs}
\label{wigner.graph.sec}

The so-called Wigner function was introduced by Wigner in order to study 
systems with a potential extending over an infinite physical space. For graphs,
we cannot use the same definition since each bond is either finite or
semi-infinite. We define a Wigner function for a graph in the following way. On
each (non-directed) bond
$a$ the Wigner function is given by 
\begin{equation}
f_{a}(x,p)=\frac{1}{2\pi \hbar }\int_{-2x}^{+2x}dy\,e^{i\frac{py}{\hbar }
}\;\psi _{a}(x-y/2)\psi _{a}^{\ast }(x+y/2)\quad \text{for}\quad 0<x<\frac{
l_{a}}{2}  \label{Wigner.graph.1}
\end{equation}
and 
\begin{equation}
f_{a}(x,p)=\frac{1}{2\pi \hbar }\int_{-(2l_{a}-2x)}^{+(2l_{a}-2x)}dy\,e^{i
\frac{py}{\hbar }}\;\psi _{a}(x-y/2)\psi _{a}^{\ast }(x+y/2)\quad \text{for}
\quad \frac{l_{a}}{2}<x<l_{a}  \label{Wigner.graph.2}
\end{equation}
In this way, the argument of the wavefunctions always remains in the interval 
$(0,l_{a})$ corresponding to the bond $a$. We notice that
$f_{a}(x=0,p)=f_{a}(x=l_{a},p)=0$ with  of the definitions
(\ref{Wigner.graph.1}) and (\ref{Wigner.graph.2}).

The Wigner function is a ``representation'' of the wavefunction in phase space
and it is essential to have a unique correspondence between the Wigner
function and the wavefunction. In order to show that this is the case with our
definition, we multiply Eq.(\ref{Wigner.graph.1}) by $e^{i\frac{py}{\hbar
}}$\ and we integrate with respect to $p$ to get 
\begin{equation}
\int_{-\infty }^{+\infty }dp\,e^{i\frac{py}{\hbar }}f_{a}(x,p)=\psi
_{a}(x+y/2)\psi _{a}^{\ast }(x-y/2)
\label{Wigner.back}
\end{equation}
If we set $y=0$ we obtain the probability density on the bond $a$
\begin{equation}
\vert \psi_{a}(x)\vert^2=\int_{-\infty }^{+\infty}dp\; f_{a}(x,p)
\label{Wigner.density}
\end{equation}
On the other hand, if we set $x=y/2$ in Eq.(\ref{Wigner.back}) we get
\begin{equation*}
\int_{-\infty }^{+\infty }dp\,e^{i\frac{py}{\hbar }}f_{a}(y/2,p)=\psi
_{a}(y)\psi _{a}^{\ast }(0)
\end{equation*}
which is valid for $0<x=y/2<\frac{l_{a}}{2}$, i.e. $0<y<l_{a}$, and we
recovered the wavefunction (or its conjugate) on all the bonds except for a
constant factor that is fixed by the boundary condition and normalization. We
could also have proceeded in a similar way with Eq.(\ref{Wigner.graph.2}).
We notice that this result shows that there is redundant information in the
Wigner function. The conclusion is that the wavefunction on all the bonds is
encoded in all the Wigner functions $\lbrace f_a(x,p)\rbrace_{a=1}^B$.

Since a wavefunction can always be written in terms of the propagator (which
is also solution of the Schr\"{o}dinger equation and represents the
evolution from a delta localized initial state) we will compute the Wigner
function for the propagator. Other cases are obtained by convenient averages
over the initial conditions. Thus we need to compute $K(x-y/2,x^{\prime
};t)K^{\ast }(x+y/2,x^{\prime };t)$. With this purpose, we use the
expression (\ref{prop.path}) which expresses the propagator as a sum over
paths (in this section we use the letter $s$ to refer to paths in order to
avoid confusion with the momentum $p$): 
\begin{equation*}
K_{a}(x-y/2,x^{\prime };t)K_{a}^{\ast }(x+y/2,x^{\prime };t)=\frac{m}{2\pi
\hbar t}\sum_{ss^{\prime }}A_{s}A_{s^{\prime }}e^{i\frac{m\left[
l_{s}(x-y/2,x^{\prime })^{2}-l_{s^{\prime }}(x+y/2,x^{\prime })^{2}\right]
}{2\hbar t}}
\end{equation*}
With $l_{s}(x,x^{\prime })$ the length of the trajectory $s$ that joins
$x^{\prime }$ to $x$.

We note that for a path that starts at $x^{\prime }$ on a bond $b_{0}$ and
ends at $x$ on the bond $b$ the lengths can be

\begin{itemize}
\item  
$l_{s}(x,x^{\prime })=(l_{b_{0}}-x^{\prime })+\widetilde{l_{s}}+x$, 
if the path goes from $x^{\prime }$ to the end of $b_{0}$ then eventually
traverses other bonds adding a distance $\widetilde{l_{s}}$ and arrives at
the position $x$ of the bond $b$ via its origin.

\item  
$l_{s}(x,x^{\prime })=x^{\prime }+\widetilde{l_{s}}+x$,
if the path goes from $x^{\prime }$ to the origin of $b_{0}$ then eventually
traverses other bonds adding a distance $\widetilde{l_{s}}$ and arrives at
the position $x$ of the bond $b$ via its origin.

\item  
$l_{s}(x,x^{\prime })=(l_{b_{0}}-x^{\prime })+\widetilde{l_{s}}+(l_{b}-x)$,
if the path goes from $x^{\prime }$ to the end of $b_{0}$ then eventually
traverses other bonds adding a distance $\widetilde{l_{s}}$ and arrives at
the position $x$ of the bond $b$ via the end of $b$.

\item  
$l_{s}(x,x^{\prime })=x^{\prime }+\widetilde{l_{s}}+(l_{b}-x)$,
if the path goes from $x^{\prime }$ to the origin of $b_{0}$ then eventually
traverses other bonds adding a distance $\widetilde{l_{s}}$ and arrives at
the position $x$ of the bond $b$ via its end.
\end{itemize}

We evaluate now the difference $l_{s}(x-y/2,x^{\prime
})-l_{s}(x+y/2,x^{\prime })$ for equal paths. We obtain 
\begin{equation*}
l_{s}(x-y/2,x^{\prime })-l_{s}(x+y/2,x^{\prime })=-y
\end{equation*}
\begin{equation*}
l_{s}(x-y/2,x^{\prime })-l_{s}(x+y/2,x^{\prime })=+y
\end{equation*}
The first result holds for trajectories that arrives at the final bond via
its origin and the second result for trajectories that arrives via the end.
Now we compute the sum which is in both cases 
\begin{equation*}
l_{s}(x-y/2,x^{\prime })+l_{s}(x+y/2,x^{\prime })=2\; l_{s}(x,x^{\prime })
\end{equation*}
Using the identity $a^{2}-b^{2}=(a+b)(a-b)$ we have the results 
\begin{equation*}
l_{s}^{2}(x-y/2,x^{\prime })-l_{s}^{2}(x+y/2,x^{\prime
})=-2\; y\; l_{s}(x,x^{\prime })
\end{equation*}
\begin{equation*}
l_{s}^{2}(x-y/2,x^{\prime })-l_{s}^{2}(x+y/2,x^{\prime
})=+2\; y \; l_{s}(x,x^{\prime })
\end{equation*}
for paths that arrives through the origin or the end of the bond $b$
respectively. We took care of equal paths because we want to separate their
contribution to the Wigner function which we call the diagonal term 
\begin{multline}
\left[ K_{a}(x-y/2,x^{\prime };t)K_{a}^{\ast }(x+y/2,x^{\prime };t)\right]
_{\rm diag}=  \label{diagonalterm} \\
\frac{m}{2\pi \hbar t}\left\{ \sum_{s_{1}}A_{s_{1}}^{2}e^{-iy\frac{
ml_{s_{1}}(x,x^{\prime })}{\hbar t}}+\sum_{s_{2}}A_{s_{2}}^{2}e^{+iy\frac{
ml_{s_{2}}(x,x^{\prime })}{\hbar t}}\right\}   \notag
\end{multline}
where $s_{1}$ is the set of paths that arrive at the bond $a$ through the
vertex at the origin of the bond $a$ and $s_{2}$ are those paths that arrive
through the vertex at the end.

For the non-diagonal term the differences in the exponent are always of the
form 
\begin{equation*}
l_{s}^{2}(x-y/2,x^{\prime })-l_{s^{\prime }}^{2}(x+y/2,x^{\prime
})=[l_{s}(x,x^{\prime })-l_{s^{\prime }}(x,x^{\prime })][l_{s}(x,x^{\prime
})+l_{s^{\prime }}(x,x^{\prime })\pm y]
\end{equation*}
or 
\begin{equation*}
l_{s}^{2}(x-y/2,x^{\prime })-l_{s^{\prime }}^{2}(x+y/2,x^{\prime
})=[l_{s}(x,x^{\prime })-l_{s^{\prime }}(x,x^{\prime })\pm
y][l_{s}(x,x^{\prime })+l_{s^{\prime }}(x,x^{\prime })]
\end{equation*}
hence we get non-diagonal terms of the form 
\begin{multline}
\lbrack K_{a}(x-y/2,x^{\prime };t)K_{a}^{\ast }(x+y/2,x^{\prime
};t)]_{\rm non-diag}=  \notag \\
\frac{m}{2\pi \hbar t}\sum_{s\neq s^{\prime }}A_{s}A_{s^{\prime }}e^{i\frac{m
\left[ l_{s}(x,x^{\prime })^{2}-l_{s^{\prime }}(x,x^{\prime })^{2}\right] }{
2\hbar t}}e^{\pm iy\frac{m\left[ l_{s}(x,x^{\prime })\pm l_{s^{\prime
}}(x,x^{\prime })\right] }{2\hbar t}}  \label{non-diagonalterm}
\end{multline}
The calculation of the Wigner function (\ref{Wigner.graph.1})-(\ref
{Wigner.graph.2}), requires the evaluation of integrals of the form 
\begin{equation}
\int_{-x_{0}}^{+x_{0}}e^{i(p\pm \Omega )\frac{y}{\hbar }}dy=2\hbar \frac{\sin 
\frac{x_{0}}{\hbar }(p\pm \Omega )}{p\pm \Omega }  \label{cuasidelta}
\end{equation}
where $\Omega =\frac{ml_{s}(x,x^{\prime })}{t}$ for the diagonal term and
$\Omega =\frac{m\left[ l_{s}(x,x^{\prime })\pm l_{s^{\prime }}(x,x^{\prime
})\right] }{2t}$ for the non-diagonal terms. In the classical limit we have
that 
\begin{equation}
\lim_{\hbar \rightarrow 0}\frac{\sin \frac{x_{0}}{\hbar }(p\pm \Omega )}{p\pm
\Omega }=\pi \delta (p\pm \Omega )
\end{equation}
from which we obtain the Wigner function in the limit $\hbar \rightarrow 0$.
In the classical limit, the phase variations of the non-diagonal terms are so
wild that the total sum is zero due to destructive interferences. We
have thus the result that, in the classical limit $\hbar\to 0$, the Wigner
function defined for graphs becomes 
\begin{equation}
f_{b}(x,p;t) \simeq \frac{m}{2\pi \hbar t}\left\{
\sum_{s_{1}}A_{s_{1}}^{2}\delta
\left[ p+ml_{s_{1}}(x)/t\right]+\sum_{s_{2}}A_{s_{2}}^{2}\delta
\left[p-ml_{s_{2}}(x)/t\right]\right\} 
\label{wigner.classical.limit}
\end{equation}
This limit corresponds to the motion of the classical density in the phase
space of the corresponding classical system. In the next section, we shall
establish that, indeed, the classical dynamics evolves the probability
density in phase space according to Eq.(\ref{wigner.classical.limit}).


\subsection{Classical dynamics on graphs}

\label{sec.class.dyn}

We have computed the classical limit for the Wigner function on graphs [see
Eq.(\ref{wigner.classical.limit})], which should be solution of the
classical ``Liouville'' equation. 
In this section, we shall summarize the main result obtained in
Ref. \cite{art4}, where we studied in detail the classical dynamics on graphs,
and we shall show that the density (\ref{wigner.classical.limit}) is 
the solution of the classical equation.
Therefore, the classical dynamics which we discuss here 
\textit{is} the classical limit of the quantum dynamics on graphs as
obtained from the classical limit of the Wigner functions.

On a graph, a particle moves freely as long as it stays on a bond. 
At the vertices, we have to introduce transition
probabilities $P_{bb^{\prime }}=\left| T_{bb^{\prime }}\right| ^{2}
$. This choice is dictated by the quantum-classical correspondence as we
shall see in this section.  The dynamics is expressed
by the following master equation (we consider the notation $x_b=[b,x]$):
\begin{equation}
\rho \left( \left[ b,x \right] ,t\right) =\sum_{b^{\prime }}P_{bb^{\prime
}}\rho \left( \left[ b^{\prime },x ^{\prime }\right] ,t-\frac{x
+l_{b^{\prime }}-x^{\prime } }{v}\right)   \label{Ev.t}
\end{equation}
The time delay correspond to the time that take to arrive
from $x'$ in the bond $b'$ to $x$ in the bond $b$.

The density $\rho \left( \left[ b,x \right] ,t\right)$
defined on each directed bond is a density defined in a
constant energy surface of phase space. In fact, the conservation of
energy fixes the modulus of the momentum and, therefore, the points
of the constant energy surface are given
by the position on the bond and the direction on the bond, that is
by position on directed bonds.

The properties of these classical dynamics for open and closed
systems are described in Ref. \cite{art4} where we also 
show that it can be understood as a random suspended flow.

To establish the connection with the classical limit of the Wigner function
we iterate the master equation (\ref{Ev.t}). In Ref. \cite{art4}
we have shown that iterating the master equation allows us to obtain
the density at the current time $t$ in terms of the density at the
initial $t_0=0$, which gives an explicit form for the Frobenius-Perron
operator 
\begin{equation}
\rho \left( \left[ b,x \right] ,t\right) =\sum_{n}\sum_{b^{\prime
}b^{\prime \prime }\cdots b^{(n)}}P_{bb^{\prime }}P_{b^{\prime }b^{\prime
\prime }}\cdots P_{b^{(n-1)}b^{(n)}}\rho \left( \left[ b^{^{(n)}},x
^{^{(n)}}\right] ,0\right)  \label{Eq.funda}
\end{equation}
with 
\begin{equation*}
x^{(n)}=x -vt+\sum_{i=1}^{n}l_{b^{(i)}}
\end{equation*}
Accordingly, $\rho \left( \left[ b,x \right] ,t\right) $ is given by a sum
over the initial conditions $\left[ b^{^{(n)}},x ^{^{(n)}}\right] $\ and
over all the paths that connect $\left[ b^{^{(n)}},x^{^{(n)}}\right] $
to $\left[ b,x \right] $ in a time $t$. Each
given path contributes to this sum by its probability
multiplied by the probability density $\rho \left( \left[
b^{^{(n)}},x^{^{(n)}}\right] ,0\right) $. 

If the initial distribution is concentrated on a point, i.e., 
\begin{equation}
\rho \left( \left[ b^{^{(n)}},x ^{^{(n)}}\right] ,0\right) =\delta
_{b^{(n)}b_{\ast }}\delta (x _{\ast }-x ^{(n)})  \label{inc.cond.dens}
\end{equation}
Eq.(\ref{Eq.funda}) can be expressed as [we use the property $\delta
(ax)=\delta (x)/a$] 
\begin{multline}
\rho \left( \left[ b,x \right] ,t\right) =\frac{1}{v}\sum_{n}\sum_{b^{\prime
}b^{\prime \prime }\cdots b^{(n)}}P_{bb^{\prime }}P_{b^{\prime }b^{\prime
\prime }}\cdots P_{b^{(n-1)}b^{(n)}}  \\
\times \delta \left( t-\frac{x
+\sum_{i=1}^{n}l_{b^{(i)}}-x_{\ast }}{v}\right) \delta _{b^{(n)}b_{\ast
}}  \label{dens.Wigner}
\end{multline}


\subsection{Connection with the classical limit of Wigner functions}

The probability density at a given time $t$ is provided by the
classical Frobenius-Perron operator (\ref{Eq.funda}). For the
particular initial condition (\ref{inc.cond.dens}) this leads to Eq.(\ref
{dens.Wigner}). Remembering that the probability of a path $s$ was written
as $A_{s}^{2}$ (we use again the letter $s$ to denote a path since $p$ is
used for the momentum) and noticing that the sum in Eq.(\ref{dens.Wigner})
is a sum over all the paths connecting $b_{\ast }$ to $b$ we can rewrite 
Eq.(\ref{dens.Wigner}) as 
\begin{equation}
\rho \left( \left[ b,x \right] ,t\right) =\frac{1}{v}\sum_{s(b_{\ast
}\rightarrow b)}A_{s}^{2}\delta \left[ t-\frac{l_{s}(x ,x_{\ast })}{v}
\right]   \label{toWigner}
\end{equation}
and by the property of the delta function 
\begin{equation*}
\frac{1}{v}\delta \left[ t-\frac{l_{s}(x ,x _{\ast })}{v}\right] =
\frac{1}{t}\delta \left[ v-\frac{l_{s}(x ,x_{\ast })}{t}\right] =
\frac{m}{t}\delta \left[ mv-\frac{ml_{s}(x ,x _{\ast })}{t}\right] 
\end{equation*}
so that Eq.(\ref{toWigner}) is equivalent to 
\begin{equation}
\rho \left( \left[ b,x \right] ,t\right) =\frac{m}{t}\sum_{s(b_{\ast
}\rightarrow b)}A_{s}^{2}\delta \left[ p-\frac{ml_{s}(x ,x _{\ast })}{t
}\right]   \label{dens.wig.}
\end{equation}
where $p=mv$. Eq.(\ref{dens.wig.}) is (up to the normalization factor $\frac{
1}{2\pi \hbar }$) the classical limit of the Wigner function as obtained in
Eq.(\ref{wigner.classical.limit}). The Wigner function 
(\ref{wigner.classical.limit}) also contains
a term with $-p$ because in Subsection 
\ref{wigner.graph.sec} we defined the Wigner
function for non-directed bonds although we deal with directed
bonds in the present section. 
Eq.(\ref{dens.wig.}) is the
probability density of being in the oriented bond $b$ with momentum $p$. In
the reverted bond $\hat{b}$ we have also a probability density which is
obtained from the density in $b$ by revering the sign of $p.$ Therefore, the
probability density of being in the non-directed bond $b$ is
\begin{equation*}
\frac{m}{t}\sum_{s_{2}(b_{\ast }\rightarrow b)}A_{s_{2}}^{2}\delta \left[ p-
\frac{ml_{s_{2}}(x ,x _{\ast })}{t}\right] +\frac{m}{t} \sum_{s_{1}(b_{\ast
}\rightarrow b)}A_{s_{1}}^{2}\delta \left[ p+\frac{ ml_{s_{1}}(x ,x _{\ast
})}{t}\right]
\end{equation*}
with $s_{1}$ and $s_{2}$ the set of paths defined for the Wigner 
function in Subsection \ref{wigner.graph.sec}.  The comparison with
Eq.(\ref{wigner.classical.limit}) shows that the quantum time evolution of the
Wigner function corresponds to the classical time evolution of the probability
density given by the classical Frobenius-Perron operator (\ref{Eq.funda}) in
the classical limit:
\begin{equation}
f_b(x,p;t) \simeq \frac{1}{2\pi\hbar} \rho([b,x],t) \, \qquad \mbox{for} \quad
\hbar\to 0
\label{semicl.Wigner.dyn}
\end{equation}
Accordingly, the classical dynamics introduced in Ref. \cite{art4}
and summarized in Subsection
\ref {sec.class.dyn} is the classical limit of the quantum dynamics on graphs
of Section \ref{sec.quan.dyn}.


\subsection{Quantum time evolution of staying probabilities}
\label{quant.evol.sec}

The preceding results shows that the classical dynamics emerges out of the
quantum dynamics of quantities such as averages or staying probabilities which
can be defined in terms of Wigner functions.  Since the Wigner functions
have a Liouvillian time evolution according to the 
Frobenius-Perron operator in the classical limit $\hbar\to 0$ we should expect
an early decay given in terms of the Pollicott-Ruelle resonances which are the
generalized eigenvalues of the Liouvillian operator.  For graphs, the
Pollicott-Ruelle resonances have been described in Ref. \cite{art4}.  Our
purpose is here to show that, indeed, the Pollicott-Ruelle resonances control
the early quantum decay of the staying probability in a finite part of an open
graph.

The quantum time evolution of the wave function is obtained from
the propagator as
$$
\pmb{\psi}(t)=\hat{\mathsf K}(t)\; \pmb{\psi}(0)
$$
where $\pmb{\psi}(0)$ represent the initial wave packet. 
As we said in Subsection \ref{propa.sec} the propagator can be obtained from
the Green function for graphs by Fourier transformation. 
Therefore we can compute $\psi_{b}(x,t)$. 

The quantum staying probability 
(or survival probability) $P(t)$ is defined as the probability of remaining in
the bounded part of an open graph at time $t$.  Since the particles that escape
cannot return to the graph, there is no recurrence and 
the staying probability is equal to the probability
of have been in the graph until the time $t$:
$$
P(t)\equiv \sum_{b}\int_{0}^{l_{b}}dx|\psi_{b}(x,t)|^2
$$
To compute $P(t)$ we can proceed as follows. First, we consider some
initial wave packet $\psi_{1}(y,0)$ on the bond 1 and with mean value $\bar E$.
The Green function $G_{b}(x,y,E)$ (where the
index $b$ refer to the coordinate $x$ on the bond $b$ and $y$ is a coordinate
on the bond 1 ) is computed
from Eqs.(\ref{green.func}) and (\ref{mu.q}). Then the propagator 
is obtained from the Fourier transform 
$$
K_{b}(x,y,t)=\int_{{\cal C}_{+}+{\cal C}_{-}}
G_{b}(x,y,E)\exp\left(-\frac{i}{\hbar}Et\right)dE
$$ 
and the wavefunction is given by
$$
\psi_{b}(x,t)=\int_{0}^{l_1}dy\,K_{b}(x,y,t)\psi_{1}(y,0)
$$

$P(t)$ is the quantum analog of the density distribution integrated over
the graph, that is the classical staying probability.
For a classically chaotic graph we know that the classical staying
probability decay
exponentially with a decay rate given by the leading Pollicott-Ruelle
resonance (see Ref. \cite{art4}).
From semiclassical arguments, the quantum staying probability $P(t)$ should
follow the classical decay for short times.  Indeed, using
Eq.(\ref{Wigner.density}), the quantum staying probability can be expressed in
terms of the Wigner function (\ref{Wigner.graph.1}) according to
\begin{equation}
P(t)=\sum_b \int_0^{l_b} dx \int_{-\infty}^{+\infty} dp f_b(x,p;t)
\end{equation}
In the classical limit $\hbar\to 0$, Eq. (\ref{semicl.Wigner.dyn}) implies
that the quantum staying probability evolves as
\begin{equation}
P(t)\simeq \sum_b \int_0^{l_b} dx \int_{-\infty}^{+\infty} dp \;
\frac{1}{2\pi\hbar} \, \rho([b,x],t)
\end{equation}
For an energy distribution well localized around the mean energy $\bar E$ of
the initial wave packet, we can suppose that the classical evolution takes
place essentially on the energy shell of energy $\bar E$.

If we denote by $\hat P^t$ the classical Frobenius-Perron operator and by
$\rho_0$ the initial probability density corresponding to the initial
wavepacket, the quantum statying probability can be written as
\begin{equation}
P(t)\simeq \langle A\vert\hat P^t\rho_0\rangle \; , \qquad \mbox{for}
\quad \hbar\to 0
\end{equation}
where we have introduced the observable $A$ defined by
\begin{eqnarray}
A[b,x] &=& \frac{l_b}{2\pi\hbar} \qquad \mbox{for}\quad b\in{\bf b} \nonumber\\
A[c,x] &=& 0 \qquad\quad \mbox{for}\quad c\in{\bf c} \nonumber
\end{eqnarray}
which is the indicator function of the bounded part of the open graph.
Using the spectral decomposition of the Frobenius-Perron operator described in
Ref. \cite{art4}, the quantum staying probability has thus the following early
decay
\begin{equation}
P(t)\simeq \sum_j \langle A\vert V_j\rangle
e^{s_jt}\langle\tilde V_j\vert\rho_0\rangle\; ,\qquad\mbox{for}\quad
\hbar\to 0
\end{equation}
in terms of the left- and right eigenvectors of the Frobenius-Perron operator:
\begin{equation}
\hat P^tV_j=e^{s_jt}V_j \; \qquad\mbox{and}\qquad \hat
P^{t\dagger}\tilde V_j=e^{s_j^*t}\tilde V_j
\end{equation}
(see Ref. \cite{art4}).
Since the leading Pollicott-Ruelle resonance is the classical escape rate for
an open graph, we can conclude that the quantum staying probability will have
an exponential early decay according to
\begin{equation}
P(t) \sim \exp\left[-\gamma_{\rm cl}(\bar v) t\right]
\end{equation}
in terms of the classical escape rate $\gamma_{\rm
cl}(\bar v)=\bar v\gamma_{\rm cl}(v=1)$ where $\bar v=\sqrt{2\bar E/m}$ is the
velocity of the classical particle at the mean energy $\bar E= m\bar v^2/2$.


\subsection{The classical zeta function and the Pollicott-Ruelle resonances}

\label{sec.P.R.res}

We have shown in Ref. \cite{art4} that he Pollicott-Ruelle resonances of a
classical particle moving with velocity $v$ on graph can be computed as the
complex zeros $\left\{s_{j}\right\}$ of the classical Selberg-Smale zeta
function of the graph given by
\begin{equation*}
Z_{\rm cl}(s) = \det \left[ {\mathsf I}-{\mathsf Q}(s)\right] =0 \; .
\end{equation*}
in terms of the matrix 
$$
Q_{bb'}(s)=P_{bb'}e^{-s\frac{l_{b'}}{v}}
$$
where $P_{bb'}=\vert T_{bb'}\vert^2$ are the transition probabilities.
This classical zeta function can be rewritten as a product over all
the prime periodic orbits on the graph as \cite{art4}:
\begin{equation}
Z_{\rm cl}(s)=\Pi _{p}\left[1-e^{-(\tilde\lambda
_{p}l_p+s\frac{l_{p}}{v})}\right]
\label{selbergsmale.graph}
\end{equation}

The zeros of the classical zeta function for a scattering system
are located in the half-plane $\mathrm{Re}\,s_{j}< 0$ 
and there is a gap empty of resonances below the axis 
$\mathrm{Re}\,s_{j}= 0$.  This gap is determined by the classical escape rate
which is the leading Pollicott-Ruelle resonance: $s_0=-\gamma_{\rm cl}$. 

\subsection{Emerging diffusion in spatially extended graphs}
\label{sec.diff}

On spatially extended graphs, the classical motion becomes diffusive.
We showed in Ref. \cite{art4} that the diffusion coefficient for 
a periodic graph can be computed from the Pollicott-Ruelle
resonances of the extended system by introducing a classical wavenumber $q$
associated with the classical probability density.  Therefore, the leading
Pollicott-Ruelle resonance acquires a dependence on this wavenumber according
to
\begin{equation}
s_0(q) = -D q^2 + {\cal O}(q^4)
\end{equation}
where $D$ is the diffusion coefficient.

If the spatially extended periodic chain is truncated to keep only $N$ unit
cells and semi-infinite leads are attached to the ends, we have furthermore
shown in Ref. \cite{art4} that the classical escape rate depends on the
diffusion coefficient according to
\begin{equation}
\gamma_{\rm cl}(N)\approx D \; \frac{\pi^2}{N^2}
\label{gamma}
\end{equation}
in the limit $N\rightarrow \infty$.

The previous results on the classical-quantum correspondence show that this
diffusive behavior is expected in the early time evolution of the quantum
staying probability for such spatially extended open graphs.  This result will
be illustrated in Section \ref{sec.ejem}.


\section{The quantum scattering resonances}

\label{sec.quantum.res}

\subsection{Scattering resonances}

\label{sec.scatt.res}

The scattering resonances are given by the poles of the
scattering matrix $\mathsf S$ in the complex plane of the quantum wavenumber
$k$. These poles are the complex zeros of 
\begin{equation*}
Z(k) \equiv \det [{\mathsf I}-{\mathsf R}(k)]
\end{equation*}
This function can be expressed as a product over periodic orbits, using the
identity $\ln \det ({\mathsf I}-{\mathsf R})={\rm tr}\ln ({\mathsf I}-{\mathsf
R})$ and the series $\ln ({\mathsf I}-{\mathsf R})=-\sum_{n\geq
1}{\mathsf R}^{n}/{n}$.  One gets 
\begin{equation}
Z(k)=\prod_{p}\left[ 1-e^{-\tilde\lambda _{p}l_{p}/2}e^{i(kl_{p}+\pi \mu
_{p}/2)}\right]  \label{zeta.Qgraphs}
\end{equation}
where $l_p$ is the length of the prime periodic orbit $p$, $\tilde\lambda$ its
Lyapunov exponent per unit length, and $\mu_p$ its Maslov index.  This formal
expression is equally valid for open and closed graphs and thus their zeros
give the eigenenergies in the first case (zeros in the real
$k$ axis) and the quantum scattering resonances in the second case, but, as
for the trace formula, the product over primitive periodic orbits does not
converge in the real axis.  

A remark is here in order about the difference between the quantum scattering
resonances and the classical Pollicott-Ruelle resonances.  The quantum
scattering resonances control the decay of the quantum wavefunction and are
defined either at complex energies $E_n$ or at complex wavenumbers or momenta
$k_n$.  In contrast, the Pollicott-Ruelle resonances control the decay of the
classical probability density which is as the square of the modulus of the
quantum wavefunction.  Accordingly, the Pollicott-Ruelle resonances are
related to the complex Bohr frequencies
$\omega_{mn}=(E_m-E_n)/\hbar$ and have the unit of the inverse of a time.

In this and the following section, we shall consider units where $\hbar=1$ and
$2m=1$, so that the quantum wavenumber $k$ is related to the energy by
$E=k^2$.  In these units, the width of a quantum scattering resonance is 
$\Gamma_n=-4\mathrm{Re}\,k_n \mathrm{Im}\,k_n$ and the
velocity of a resonance is $v_n=2\mathrm{Re}\,k_n$,
so that $\Gamma_n=-2v_n \mathrm{Im}\,k_n$.


\subsection{Topological pressure and the gap for quantum scattering
resonances}
\label{press.sec}

The chaotic properties of the classical dynamics
can be characterized by quantities such as the
topological entropy, the Kolmogorov-Sinai entropy, 
the mean Lyapunov exponent, or the partial Hausdorff dimension $d_{\rm H}$ in
the case of open systems.  All these quantities can be derived from
the so-called ``topological pressure" per unit time $P(\beta;v)$ which we
analyzed in Ref. \cite{art4} for graphs.

Beside these important properties, the topological pressure
also provides information for the quantum scattering problem. 
In fact, the quantum zeta function has a structure
very similar to a Ruelle zeta function with some exponent $\beta=1/2$
(see Refs. \cite{art4, GaspRice}).  As a consequence, the quantum zeta
function is known to be holomorphic for
$\mathrm{Im}\, k>\tilde P(\beta)$ where $\tilde P(\beta)$ is the so-called
topological pressure per unit length.  Therefore the poles of the zeta
function are located in the half plane
$${\rm Im}\; k_n \leq  \frac{1}{v_n}\;
 P\left(\frac{1}{2};v_n\right) \equiv \tilde P\left(\frac{1}{2}\right)$$

The following result can thus be deduced
\cite{GaspRice,GaspVarenna,GAB,GB,GaspLeiden}:

{\it
If $\tilde P\left(\frac{1}{2}\right) <0$ or equivalently if $0\leq d_{\rm
H}<\frac{1}{2}$, the
lifetimes $\lbrace\tau_n\rbrace$ are smaller than a maximum quantum
lifetime $\tau_{\rm q}$ and there
is a gap in the resonance spectrum.

If $\tilde P\left(\frac{1}{2}\right) \geq 0$ or equivalently if
$\frac{1}{2} \leq d_{\rm H}\leq 1$, the
lifetimes may be arbitrarily long.}

In the first case, the partial Hausdorff dimension is small ($0\leq d_{\rm
H}<\frac{1}{2}$) and we can talk about a filamentary set of trapped
trajectories.  In the second case with $\frac{1}{2} \leq d_{\rm H}\leq 1$, the
set of trapped trajectories is bulky.  Hence, the result shows that a gap
appears in the distribution of quantum scattering resonances in the case of a
filamentary set of trapped trajectories.  The gap is determined by the
topological pressure at $\beta=1/2$ which is the exponent corresponding to
quantum mechanics, as opposed to the exponent $\beta=1$ which corresponds to
classical mechanics \cite{GaspRice,GaspLeiden}.

The properties of the topological pressure yield the following important
inequality between the quantum lifetime $\tau_{\rm q}$ and the classical
lifetime $\tau_{\rm cl}=1/\gamma_{\rm cl}$:
 $$\frac{1}{\tau_{\rm q}}=-2 \;
 P\left(\frac{1}{2};v\right)
\leq  - P(1;v) =\frac{1}{\tau_{\rm cl}}$$
where the equality stands only for a set of trapped trajectories which reduces
to a single periodic orbit \cite{GaspRice,GaspVarenna}.
Accordingly, the quantum lifetime equals the classical lifetime only for a
periodic set of trapped trajectories.  On the other hand, the quantum lifetime
is longer than the classical lifetime for a chaotic set of trapped
trajectories.  This result -- which has previously been proved for billiards
and general Hamiltonian systems in the semiclassical
limit \cite{GaspRice,GaspVarenna,GAB,GB,GaspLeiden} --  thus also extends to
open graphs.

We note that it is thanks to the time continuous classical dynamics
that we can compute the above estimate of the gap of resonances.
Examples of this will be considered in Section \ref{sec.ejem}.


\subsection{Mean motion and the density of resonances}
\label{mean.motion.subsec}

Beside the possibility of a gap in the distribution of the scattering
resonances, we also want to obtain the distribution of the imaginary parts
which give the widths of the scattering resonances.  For this purpose,
we first determine the mean density of resonances. This can be obtained
analytically for a general graph and was done by Kottos and Smilansky who
obtained a trace formula for the resonance density. Here, we proceed in a
different way. The zeta function for a $k$-independent matrix
$T$ (e.g. with Neumann boundary conditions), is an almost-periodic function of
$k$ and several results are known about their properties, in particular:

The mean density of resonances $H(y_{1},y_{2})$ (or zeros of the zeta
function) in a strip $y_{1}<\mathrm{Im}\,k<y_{2}$ of the complex plane
$k=x-iy$ is determined from the number of resonances
$N(x_{1},x_{2},y_{1},y_{2})$ in the rectangle $(x_{1},x_{2},y_{1},y_{2})$ by
the following relation \cite{almost.periodic} 
\begin{equation*}
H(y_{1},y_{2})=\lim_{|x_{2}-x_{1}|\rightarrow \infty
}\frac{N(x_{1},x_{2},y_{1},y_{2})}{x_{2}-x_{1}}=\frac{1}{2\pi }\left[
M(y_{2})-M(y_{1})\right]
\end{equation*}
where $M(y)$ is the mean motion of the function of $x$ 
\begin{equation}
f_{y}(x)=Z(k=x-iy).  \label{f_y(x)}
\end{equation}

The function 
\begin{equation}
h(y)=H(0,y)=\frac{1}{2\pi }\left[ M(y)-M(0^{+})\right]
\label{den.res.meanmotion}
\end{equation}
gives the density of resonances with $\mathrm{Im}\,k<y$. The total density
of resonances is therefore given by $h(\infty )$. We shall compute this
number using the general properties of the function $Z(k)$.

\subsubsection{Mean motion}

An almost-periodic function $f:t \in  R \rightarrow C$ 
\begin{equation*}
f(t)=r(t)e^{i\phi (t)}
\end{equation*}
with real $r(t)$ and $\phi (t)$, has a mean motion $M$ 
\begin{equation*}
M=\lim_{t\rightarrow \infty }\frac{\phi (t)}{t}
\end{equation*}
if the limit exists.

The problem of computing the mean motion has a long history and is a
difficult problem. It was posed by Lagrange in $1781$ and there are still
only a few general results. For an almost-periodic function formed by three
frequencies there is a explicit formula given by Bohl. 
Weyl proved the existence
of mean motion for functions with a finite number of incommensurate
frequencies and also gave a formula to compute it.

The simplest result holds for the so-called Lagrangian case considered by
Lagrange in his original work. We quote the result because it is important
for what follows.

Consider a function 
\begin{equation}
f(t)=a_{0}e^{i\omega _{0}t}+a_{1}e^{i\omega _{1}t}+\cdots +a_{n}e^{i\omega
_{n}t}.  \label{exp.ft}
\end{equation}
If 
\begin{equation*}
|a_{0}|\geq |a_{1}|+|a_{2}|+\cdots +|a_{n}|
\end{equation*}
then $f$ has a mean motion $M$ which is $M=\omega _{0}$. We shall show that
the density of resonances is determined by the mean motion of the function
in Eq.(\ref{f_y(x)}) in the Lagrangian case.

\subsubsection{The density of resonances $h(\infty )$}

Let us consider the expansion of the determinant involved in the zeta
function [see Eq.(\ref{quant1})]. If $2B$ is the dimension of the square
matrix $\mathsf R$, then 
\begin{equation*}
\det (\lambda {\mathsf I}-{\mathsf R})=\sum_{l=0}^{2B}m_{2B-l}\lambda ^{l}
\end{equation*}
where $m_{0}=1,m_{1}={\rm tr}(-{\mathsf R})$ ... and $m_{2B}=\det (-{\mathsf
R})$. The secular equation is $\sum_{l=0}^{2B}m_{l}=0$.\ The general term 
\begin{equation*}
m_{p}=\sum_{1\leq i_{1}\leq i_{2}\leq \cdots \leq i_{p}\leq 2B}(-{\mathsf
R})\left( 
\begin{array}{ccc}
i_{1} & \cdots  & i_{p} \\ 
i_{1} & \cdots  & i_{p}
\end{array}
\right) ,
\end{equation*}
where $(-{\mathsf R})\left( 
\begin{array}{ccc}
i_{1} & \cdots  & i_{p} \\ 
i_{1} & \cdots  & i_{p}
\end{array}
\right) $ is the principal minor of order $p$ obtained by eliminating the
$n-p$ rows and columns of $(-{\mathsf R})$ different from $i_{1}\cdots
i_{p}$.  This coefficient $m_{p}$ is the homogeneous symmetric polynomial of
degree $p$ which can be constructed from the $2B$ eigenvalues of ${\mathsf
R}$, therefore $m_{p}$ is of the form\footnote{ Remember that ${\mathsf
R}(k)={\mathsf T}{\mathsf D}(k)$ according to Eq.(\ref{quant1a}).} 
\begin{equation*}
m_{p}=\sum_{\{J_{p}\}}a_{J_{p}}e^{ik\sum_{j\in J_{p}}l_{j}}\; ,
\end{equation*}
where $J_{p}$ stands for a set of $p$ different integers in the interval $
[1,2B]$ and $\{J_{p}\}$ the set of elements $J_{p}$. The important point to
notice is that in $m_{p}$ there are $p$ lengths.

It is clear that the only term which involves all the lengths of the graph
in the expansion 
\begin{equation}
f_{y}(x)=\det \left[{\mathsf I}-{\mathsf R}(x-iy)\right]=1+m_{1}+\cdots
+m_{p}+\cdots +m_{2B-1}+m_{2B}
\label{f.expan}
\end{equation}
is 
\begin{equation}
m_{2B}=\det {\mathsf R}=e^{2iL_{\rm tot}k}\det {\mathsf T}=e^{2iL_{\rm
tot}x}e^{2L_{\rm tot}y}\det {\mathsf T}
\label{star}
\end{equation}
For $y\rightarrow \infty ,$ $m_{2B}$ is exponentially larger than the
remaining terms in Eq.(\ref{f.expan})\ and the next leading term $m_{2B-1},$
which can be estimated as 
\begin{equation}
m_{2B-1}\sim e^{i(2L_{\rm tot}-l_{\rm min})k}\frac{\det T}{\lambda _{\rm min}}
\label{m2B-1}
\end{equation}
with $\lambda _{\rm min}$ the minimum eigenvalue of $\mathsf T$, is
exponentially larger than $m_{p}$ with $p=1,...,2B-2$. Therefore we have 
\begin{equation*}
\left| m_{2B}\right| >\left| m_{2B-1}\right| +\cdots +\left| 1\right| \qquad 
\text{for}\quad y\rightarrow \infty
\end{equation*}

Thus, the function $f_{y}(x)$ is in the Lagrangian case. The term $m_{2B}$
corresponds to the term $a_{0}e^{i\omega _{0}t}$ (here $t$ is replaced by $x$)
of the expansion Eq.(\ref{exp.ft}) and the mean motion is given by the
frequency of $m_{2B}$ given by Eq.(\ref{star}). Accordingly, we find that
$M(\infty)=2L_{\rm tot}$.

In order to compute the density of resonances $h(\infty )$ [see Eq.(\ref
{den.res.meanmotion})] we have to evaluate $M(0^{+})$. Since the upper half
of the complex plane is empty of resonances we have that $M(0^{+})=0$ and,
thus, from Eq.(\ref{den.res.meanmotion}) we get the density of resonances 
\begin{equation}
h(\infty )=\frac{L_{\rm tot}}{\pi }
\label{dens.tot}
\end{equation}
This result prevails as long as $\det {\mathsf T}\neq 0$. In the opposite case
the density is given by 
\begin{equation}
h(\infty )=\frac{2L_{\rm tot}-l_{\rm min}}{2\pi }
\label{dens.nearly.tot}
\end{equation}
if the corresponding constant factor does not vanish.

From this argument, it is clear that, once the function $f_{y}$ belongs to
the Lagrangian case (that is, when $y=y_{\max }$), no further resonance
appears below $y_{\max }$. This allows us to estimate how deep in the
complex plane lies the shortest living resonances. With
this aim, we consider the
largest terms of the expansion (\ref{f.expan}), i.e., 
\begin{equation*}
f_{y}(x)=m_{2B}+m_{2B-1}+\cdots 
\end{equation*}
with $m_{2B}=e^{2iL_{\rm tot}k}\det T$ and $m_{2B-1}\sim e^{i(2L_{\rm tot}
-l_{\rm min})k} \frac{\det {\mathsf T}}{\lambda _{\rm min}}$. Thus the
Lagrangian case holds approximately when 
\begin{equation*}
e^{2L_{\rm tot}y}>e^{(2L_{\rm tot}-l_{\rm min})y}\frac{1}{\lambda _{\rm min}}
\end{equation*}
that is for 
\begin{equation}
y_{\rm max}\sim \frac{1}{l_{\rm min}}\ln \left( \frac{1}{|\lambda _{\rm min}|}\right) 
\label{y_max}
\end{equation}
This estimate turns out to be quite good as we shall see.

This result can be obtained also from the following argument. The largest
$y=|{\rm Im}\; k|$ that can be solution of Eq.(\ref{quant1}) is approximately
given by the equation 
\begin{equation*}
1-e^{ixl_{\rm min}}e^{y_{\rm max}l_{\rm min}}\lambda _{\rm min}=0
\end{equation*}
from where we obtain the result of Eq.(\ref{y_max}).

A similar argument was used by Kottos and Smilansky 
\cite{Smilansky3} to obtain the gap empty of resonances $0<y<y_{\rm min}$ with 
\begin{equation*}
y_{\rm min}\sim\frac{1}{l_{\rm max}}\ln \left( \frac{1}{|\lambda _{\rm max}|}
\right).
\end{equation*}

In Subsection \ref{press.sec}, we presented a lower bound for
this gap which is very accurate.  Chaotic systems with a fractal
set of trapped trajectories of partial Hausdorff dimension $d_{H}<1/2$, have a
gap empty of resonances below the axis $\mathrm{Re}\,k$ given by
$$y_{\rm min}=-\tilde P(1/2)$$
This bound is based on the classical dynamics.  In this case, the cumulative
function $h(y)$ vanishes for $0<y<y_{\rm min}=-\tilde P(1/2)$.

The existence of the function $h(y)$ in the limit $x={\rm Re}\; k\to\infty$ for
the case of graphs and the relations (\ref{dens.tot})-(\ref{dens.nearly.tot})
are compatible with a conjecture by Sj\"ostrand \cite{Sjostrand} and Zworski
\cite{Zworski,LZ} that the distribution of scattering resonances should obey a
generalized Weyl law expressed in terms of the Minkowski dimension of the set
of trapped trajectories, because this Minkowski dimension is equal to one for
the quantum graphs.

\subsubsection{Width's distribution}

The density of resonances with a given
imaginary part, $P(y)$, is defined by 
\begin{equation}
P(y)dy=\frac{\left\{ \#\ \text{ of resonances}\,k_{n}=x_{n}-iy_{n}\,\,
\hbox{such that}\,\,y<y_{n}<y+dy\right\} }{\left\{ \text{total }\#\ \text{ of
resonances}
\right\} }
\label{dfn.P(y)}
\end{equation}
or in terms of the previously defined $h(y)$: 
\begin{equation*}
P(y)=\frac{dh(y)}{dy}\; .
\end{equation*}

The power law $P(y)\sim y^{-3/2}$ has been
conjectured in Ref. \cite{Guarnieri} to be a generic feature of the density of
resonances for systems with diffusive classical dynamics.
The system studied in Ref. \cite{Guarnieri} was the quantum kicked rotor 
whose classical limit is the standard map.
For some values of the parameters that
appear in this map, the phase space is filled with chaotic trajectories and
not large 
quasi-periodic island are observed. Working with these particular values,
the standard map produces a deterministic diffusion. In Ref. \cite{Guarnieri},
the kicked rotor has been turned
into an open system by introducing absorbing boundary conditions at some
fixed values, say $+N/2$ and $-N/2$. 
The distribution $P(y)$ for the scattering resonances
obtained in Ref. \cite{Guarnieri}
is not identical to the one we observe for our graph.  In particular, the
density $P(y)$ of Ref. \cite{Guarnieri} starts with $P(y=0)=0$ and grow until a
maximum value $y \sim \gamma _{\rm cl}$, with $\gamma _{\rm cl}$ 
the classical escape rate.  It
is the tail of $P(y)$ which decreases from $y \approx \gamma_{\rm cl}$ 
as the power law $P(y)\sim y^{-3/2}$.

The conjecture is that the power law holds for the tail of $P(y)$,
that is for large $y$.
We present here the argument of Ref. \cite{Guarnieri} which motivates this
conjecture for systems with a diffusive classical limit.

Consider a classical open diffusive system that extends from $x=-N/2$ to
$x=+N/2$. We set absorbing boundary conditions at $x=-N/2$ and $x=+N/2$.
Consider at $t=0$ a particle in the interval $(-N/2,+N/2)$. Since the particle
escapes by a diffusion process the mean time that the particle takes to arrive
at the border, starting at a distance $X$ from it, is the diffusion time
$t_{\rm d}\sim \frac{X^{2}}{D}$.  We suppose that the resonant states are more
or less uniformly distributed along the chain and that their quantum lifetime
is proportional to the mean time taken by the particle to move from
$X$ to the border so that 
$|\mathrm{Im}\,k|=Y \sim \frac{1}{t_{\rm d}} \sim \frac{D}{X^2}$.
The chain being symmetrical under $x\to -x$, we can assume a uniform
distribution of $X$ from the border $x=-N/2$ to the middle $x=+N/2$. The
probability that the imaginary part is smaller that $y$ is thus
\begin{equation*}
{\rm Prob}\left\{Y<y\right\}={\rm
Prob}\left\{\frac{aD}{x^{2}}<y\right\}={\rm
Prob}\left\{\sqrt{\frac{aD}{y}}<X\right\} =1-\frac{2}{N}\sqrt{\frac{aD}{y}}
\end{equation*}
where $a$ is a dimensionless constant.
Therefore, the probability density of the imaginary parts of the resonances
$k_n=x_n-iy_n$ is given by
\begin{equation}
P(y )=\frac{d}{dy}{\rm Prob}\left\{Y<y\right\}=\frac{\sqrt{aD}}{N}y ^{-3/2}\; .
\label{res.dens.Guarnieri}
\end{equation}

This qualitative argument can be criticized on various points. 
In particular, the constant factor $a$ is not determined and the assumption 
that the result holds for the quantum case is questionable because of the use
of classical considerations. A complete theoretical validation of this law is
thus lacking. However, the numerical results  presented below give a support to
this conjecture in the case of a multiconnected graph which is spatially
extended (see Subsection \ref{linear.graph}).

\section{Examples and discussion}
\label{sec.ejem}

\subsection{Simple graphs with two leads}

For the first two examples in this section, we consider
Neumann boundary condition 
$\sigma _{ab}^{i}=\frac{2}{\nu_{i}}-\delta _{ab}$.

\begin{figure}[ht]
\centering
\includegraphics[clip=true,width=6cm]{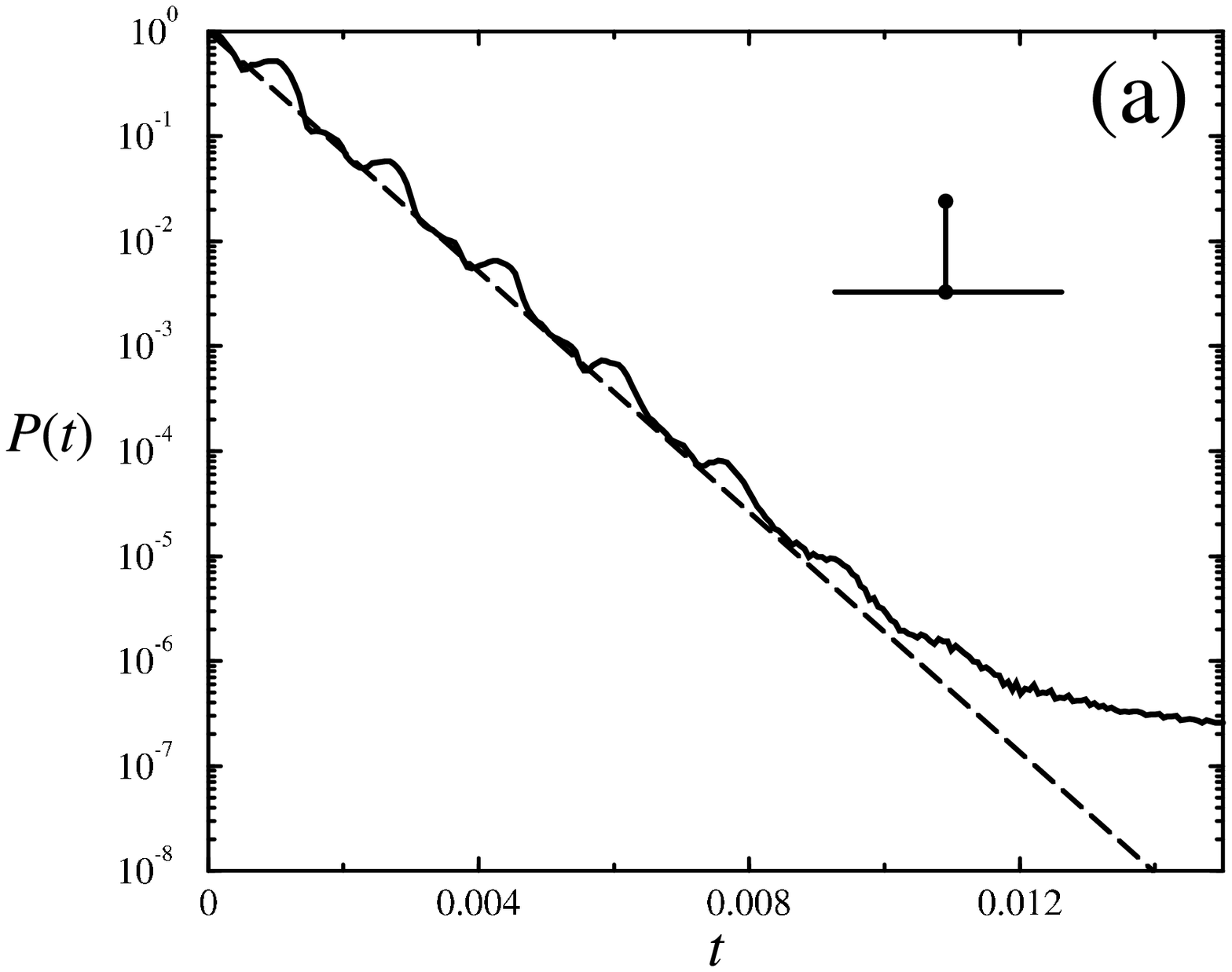}
\includegraphics[clip=true,width=6cm]{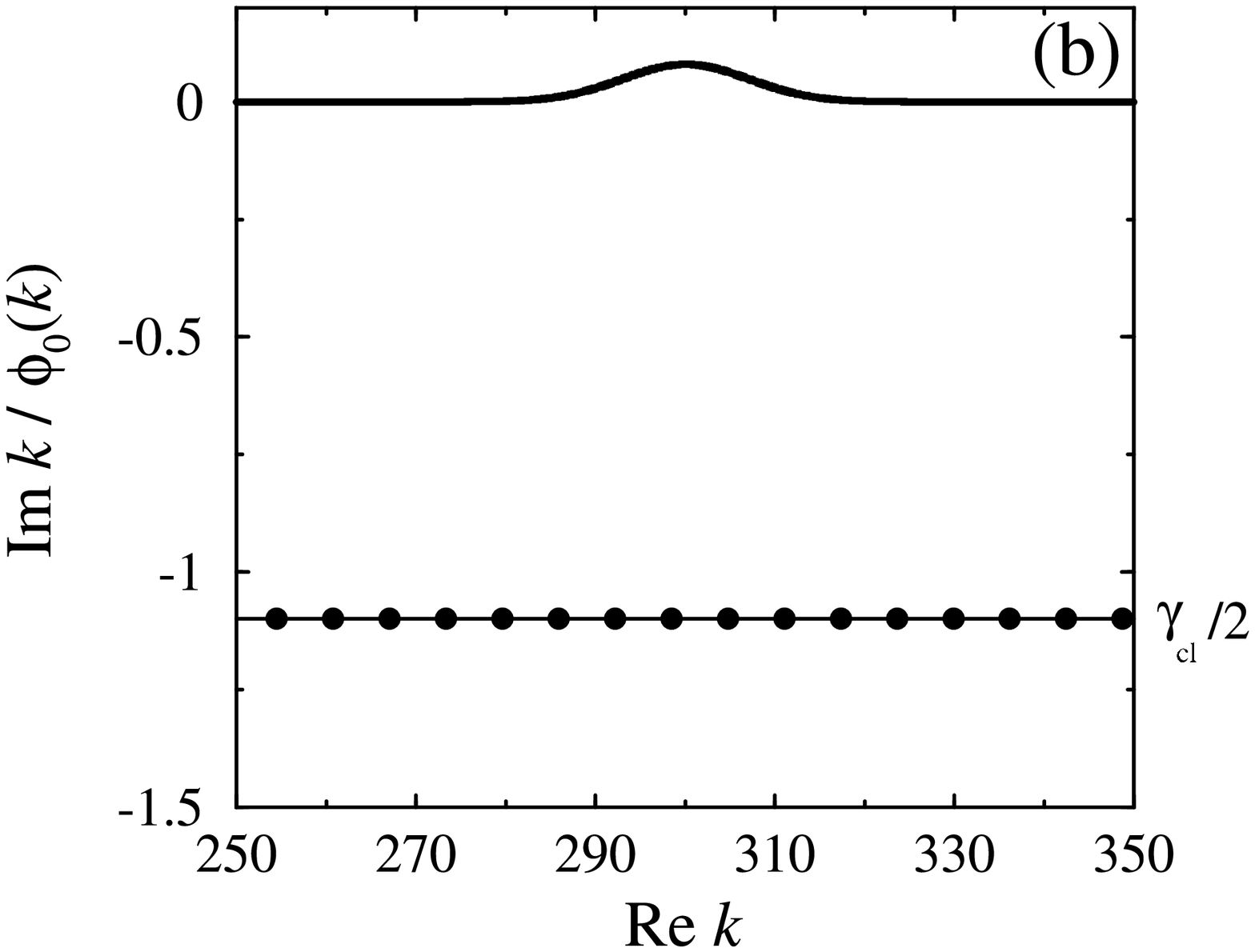}
\vskip 0.2cm
\includegraphics[clip=true,width=6cm]{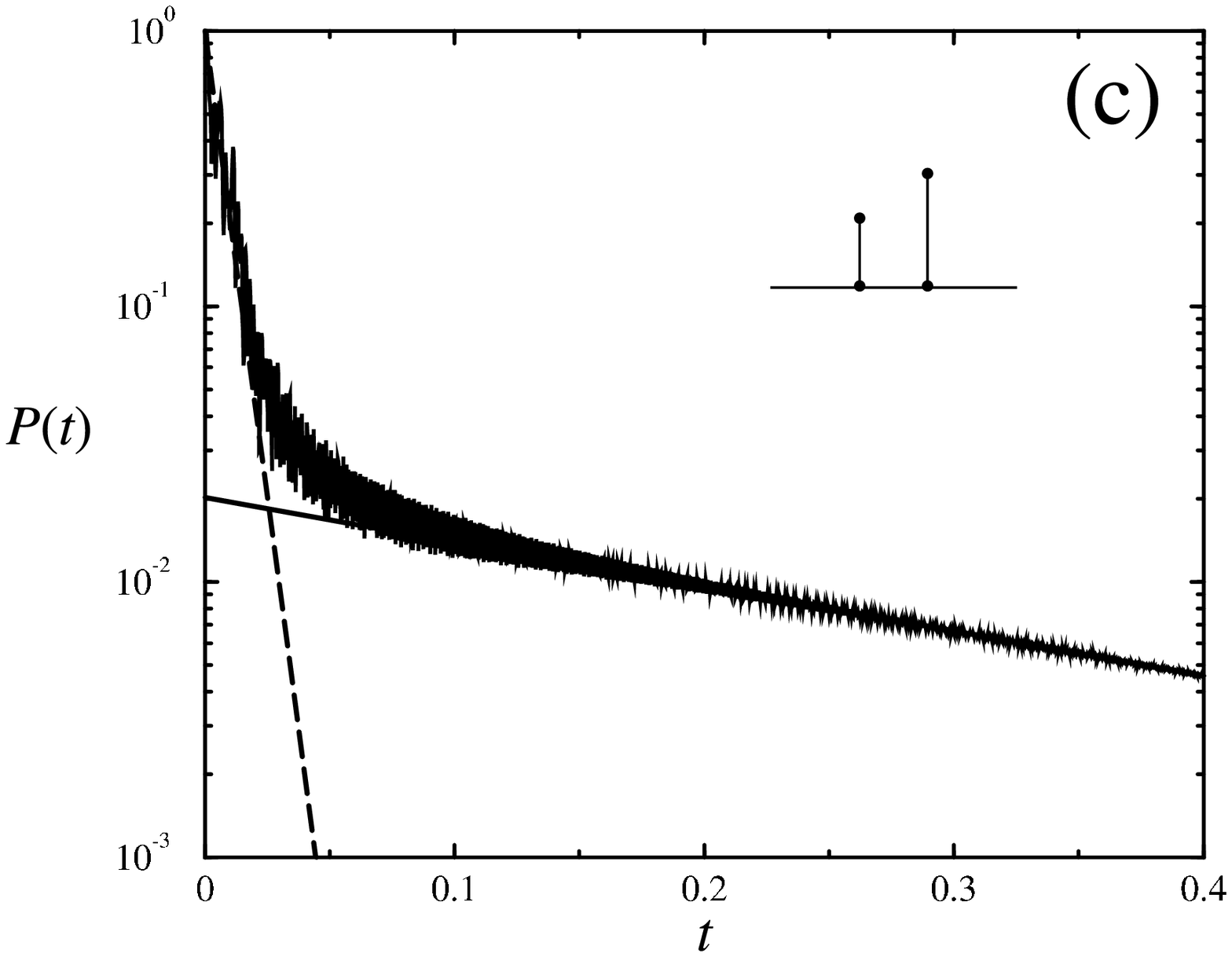}
\includegraphics[clip=true,width=6cm]{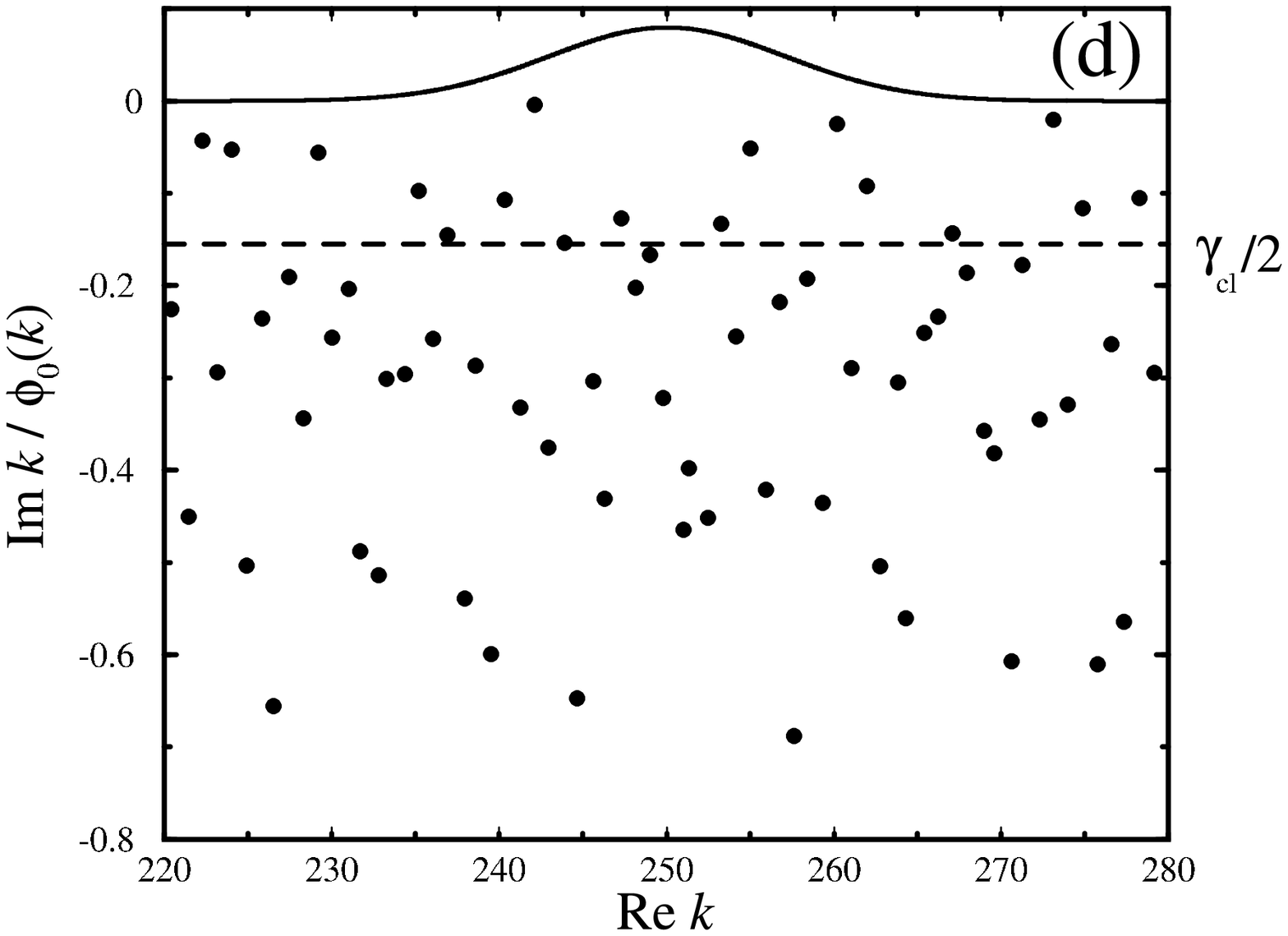}
\caption{(a) Decay of the quantum staying probability
for the graph of the inset. The straight line is the exponential decay
of the classical density for the same graph.
(b) The dots represent the quantum scattering resonances of
the graph in figure (a). The line superposed to the dots
is the corresponding value associated with the classical decay
rates of the classical Pollicott-Ruelle resonances. The curved
line on the real axis is the initial wave packet.
(c) The same as in (a) for the graph of the inset in (c).
The steepest slope corresponds to the classical decay
while the line with the smaller slope corresponds
to the decay determined from the isolated resonance which is the closest to
the real axis in figure (d). (d) The same as (b) for the graph of the inset
in (c). }
\label{decay1.fig}
\end{figure}

In Figs. \ref{decay1.fig}a and \ref{decay1.fig}c,
we compare the decay of the quantum staying probability
with the classical decay obtained from the Pollicott-Ruelle resonances
and we also depict in Figs. \ref{decay1.fig}b and \ref{decay1.fig}d
the spectrum of quantum scattering resonances for the 
corresponding graph. The initial wave packet 
also plotted in Figs. \ref{decay1.fig}b and \ref{decay1.fig}d define a
spectral window.

The excellent agreement in Fig. \ref{decay1.fig}a is
due to the fact that the quantum and classical lifetimes of 
the resonances coincide. 

In fact, for the simple graph (inset Fig. \ref{decay1.fig}a),
the scattering matrix is
\begin{equation}
{\mathsf S}=\left[ 
\begin{array}{cc}
-\frac{1}{3} & \frac{2}{3} \\ 
\frac{2}{3} & -\frac{1}{3}
\end{array}
\right] +\frac{4e^{2ikg}}{9+3e^{2ikg}}\left[ 
\begin{array}{cc}
1 & 1 \\ 
1 & 1
\end{array}
\right]  \label{U.ejem}
\end{equation}
with $l_{b}=g.$ 

The resonances are given by the poles of $\mathsf S$ in Eq.(\ref{U.ejem}),
which are the zeros of $9+3e^{2ikg}$ that is 
\begin{equation}
k_{n}=\pm \frac{2n+1}{2g}\pi -i\frac{\ln 3}{2g}  \label{ex.res}
\end{equation}
with $n$ integer. These resonances have a lifetime $\tau _{n}=\frac{1}{\Gamma
_{n}}$ with 
$\Gamma _{n}=-2v_n\mathrm{Im}\,k_n=v\frac{\ln 3}{g}$
where we used $v=2\mathrm{Re}\,k$. This means that all the
resonances have the lifetime given by 
\begin{equation}
\tau _{n}=\frac{g}{v\ln 3}\; .  \label{ex.quantum.life}
\end{equation}
This result could have been obtained using Eq.(\ref{zeta.Qgraphs})
and the classical Pollicott-Ruelle resonances from 
Eq.(\ref{selbergsmale.graph}). 

In fact, for this graph, the only periodic orbit is the one that bounces 
on the bond $b$. The length of this periodic orbit is $l_{p}=2g$.
The stability coefficient is given by $|A_{p}|^{2}=\exp (-\tilde\lambda _{p}l_{p})=
\left[ \sigma _{\hat{b}b}^{2}\sigma _{b\hat{b}}^{1}\right] 
=1\times \frac{1}{9}$ and we obtain $\tilde\lambda _{p}l_{p}=\ln 9.$
At the vertex $2$ the particle
is reflected with trivial back scattering and thus the analogue of the
Maslov index $\mu _{p}=2$ for the periodic orbit.   
Thus the quantum scattering resonances are the solutions of 
\begin{equation*}
1-\exp \left( \frac{-\ln 9}{2}+2igk+i\pi \right) =0
\end{equation*}
from which Eq.(\ref{ex.res}) follows, while the classical Pollicott-Ruelle
resonances are given by the solutions of
\begin{equation*}
1-\exp \left(\ln 9-2gs/v\right) =0
\end{equation*}
which follows from Eq.(\ref{selbergsmale.graph}). These solutions are
\begin{equation}
s_{n}=-\frac{\ln 9}{2g/v}\pm \frac{2in\pi }{2g/v}
\end{equation}
Therefore, all the Pollicott-Ruelle 
resonances have the lifetime $\tau _{\rm cl}=\frac{2g}{v\ln 9}=\frac{g}{v\ln
3}$. This lifetime coincides with the quantum lifetime obtained from the
resonances of the same graph in Eq.(\ref{ex.quantum.life}). 

The second graph in Figs. \ref{decay1.fig}c and \ref{decay1.fig}d
has a chaotic classical dynamics.  In this case, we also observe a very good 
agreement for short times between the quantum staying probability and the
classical prediction and then a transition to a pure quantum regime. The
decay in the quantum regime is again exponential because there
is an isolated resonance that controls the long-time decay. 
In fact, as we see in Fig. \ref{decay1.fig}d, there is an isolated resonance
very near the real axis under the window given by the initial wave packet.
The decay rate given by this resonance,
$\Gamma_n=-4 {\mathrm Re}\,k_n {\mathrm Im}\,k_n$,
gives the straight line with the smaller slope.

\subsection{Triangular graphs}

\begin{figure}[ht]
\centering
\includegraphics[width=8cm]{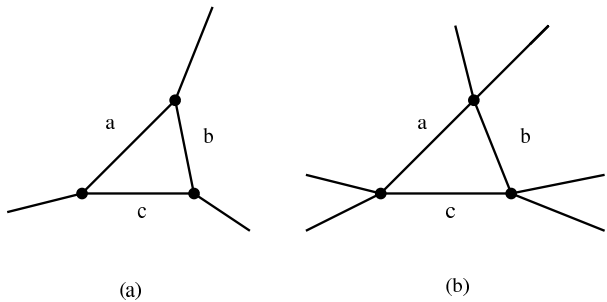}
\caption{The open triangle quantum graph: (a) One scattering
lead is connected to each vertex. (b) Two scattering leads are
connected to each vertex.}
\label{fig.triang}
\end{figure}

The following example is a nice illustration of the role played by the
Lagrangian mean motion in the density of resonances described in Subsection
\ref{mean.motion.subsec}. Consider a graph with the form of a triangle, that
is we have three vertices and every vertex is connected to the two other
vertices. The length of the bonds are
$a$, $b$ and $c$. See Fig. \ref{fig.triang}.

Now, we add a semi-infinite scattering lead to each vertex and we use Neumann
boundary condition at the vertices (see Fig. \ref{fig.triang}a). 
In this case, the resonances are determined by the zeros of
\begin{multline*}
f(k)=e^{2ik(a+b+c)}-16e^{ik(a+b+c)}-e^{2ik(a+b)}-e^{2ik(b+c)} \\
-e^{2ik(a+c)}-3(e^{ika}+e^{ikb}+e^{ikc})+27=0
\end{multline*}
replacing $k=x-iy$ we have that the mean motion of $f_{y}(x)$ for
$y\rightarrow \infty $ is $M(y\rightarrow \infty )=2(a+b+c)$ and the density
is given by $h(\infty )=\frac{a+b+c}{\pi }$, which illustrates Eq.
(\ref{dens.tot}).

Now, consider the same graph but with two semi-infinite leads attached to
each vertex (see Fig. \ref{fig.triang}b). In this case, the resonances are determined by the equation 
\begin{equation*}
e^{2ika}+e^{2ikb}+e^{2ikc}+e^{ik(a+b+c)}=4
\end{equation*}
replacing $k=x-iy$ we have that the mean motion of $f_{y}(x)$ for
$y\rightarrow \infty $ is given by $\max \{a+b+c,2a,2b,2c\}$ and therefore the
density of resonances is given by $h(\infty )=\max \{\frac{a+b+c}{2\pi
},\frac{a}{\pi },\frac{b}{\pi },\frac{c}{\pi }\}$, which illustrates Eq.
(\ref{dens.nearly.tot}).

An interesting observation is the following. The matrix $\mathsf T$ that
contains the transmission and reflection amplitudes for the triangular graph
can be written as 
\begin{equation*}
{\mathsf T}_{\beta }=\left[ 
\begin{array}{cccccc}
0 & 0 & u & 0 & v & 0 \\ 
0 & 0 & v & 0 & u & 0 \\ 
u & 0 & 0 & 0 & 0 & v \\ 
v & 0 & 0 & 0 & 0 & u \\ 
0 & u & 0 & v & 0 & 0 \\ 
0 & v & 0 & u & 0 & 0
\end{array}
\right]
\end{equation*}
with $u=-\frac{\beta }{2}-\frac{(1-\beta )}{3}$ and $v=\frac{\beta
}{2}+\frac{2(1-\beta )}{3}$ with $\beta =1$ for the graph connected to two
scattering leads at each vertex and $\beta =0$ when there is only one
scattering lead per vertex. Note that $\det {\mathsf T}=0$ if $\beta =1$. We
have computed the zeros of the function $Z_{\beta }(k)=\det [{\mathsf
I}-{\mathsf T}_{\beta }{\mathsf D}(k)]$ for $0\leq \beta \leq 1$.  It is
observed that some zeros in the lower half of the complex $k$ plane decrease
their imaginary part very fast as $\beta$ increases. Therefore, we interpret
the lowering of the density for the triangle connected to two scattering
leads per vertex as the effect of resonances with $\left|
\mathrm{Im}\,k\right| \rightarrow \infty$.  However, it should be noted that,
even in the case that $\det {\mathsf T}=0$, the analysis leading to Eq.
(\ref{y_max}) (i.e., the competition between the two leading terms of the
function $f_{y}(x)$ for $y\rightarrow \infty $) permits to obtain an
approximate upper bound $y_{\rm max}$.


\subsection{Spatially extended multiconnected graph}

\begin{figure}[th]
\centering
\includegraphics[width=8cm]{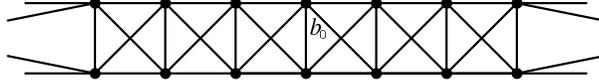}
\caption{Open graph built out of a periodic chain by attaching semi-infinite
leads on a graph made of $N$ unit cells. The figure shows a chain with
$N=6$ unit cells.}
\label{fig.chain}
\end{figure}

Here, we consider the multiconnected graph of Fig. \ref
{fig.chain}. The classical dynamics on this graph
was studied in Ref. \cite{art4} where we showed that the escape is 
controlled by diffusion.

\begin{figure}[h]
\centering
\includegraphics[width=8cm]{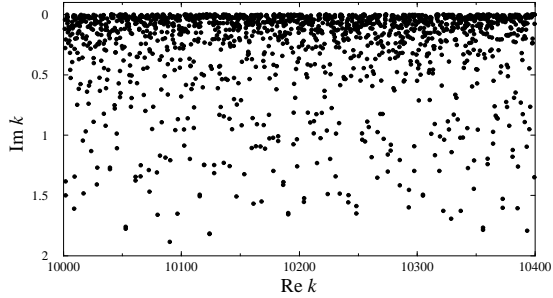}
\caption{Quantum scattering resonances for the graph in Fig. 
\ref{fig.chain} with $N=5$ unit cells.}
\label{fig.res12}
\end{figure}

For this particular graph the $\mathsf T$ matrix is 
\begin{equation*}
T_{bb^{\prime }}=\left\{ 
\begin{array}{c}
-\frac{3}{5}\text{ if the particle is reflected, i.e., }b=b^{\prime }; \\ 
\frac{2}{5}\text{ for bonds }b\neq b^{\prime }\text{which are connected;} \\ 
0\text{ otherwise.}
\end{array}
\right. 
\end{equation*}
The spectrum of quantum scattering resonances of this graph is depicted in
Fig. \ref{fig.res12} for the chain with $N=5$ unit cells. 

\subsubsection{Width's distribution}

\begin{figure}[tbp]
\includegraphics[width=6cm]{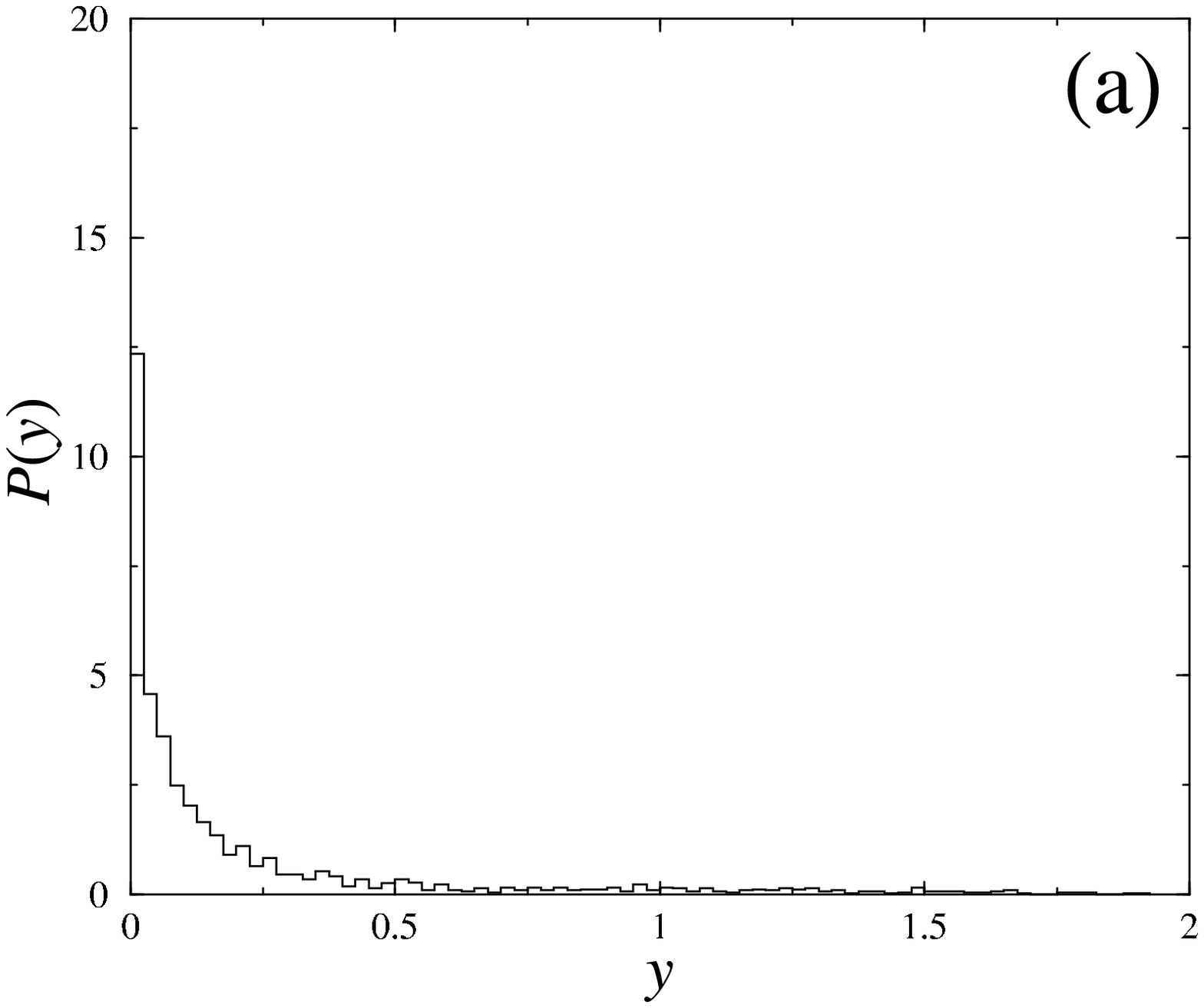}
\includegraphics[width=6cm]{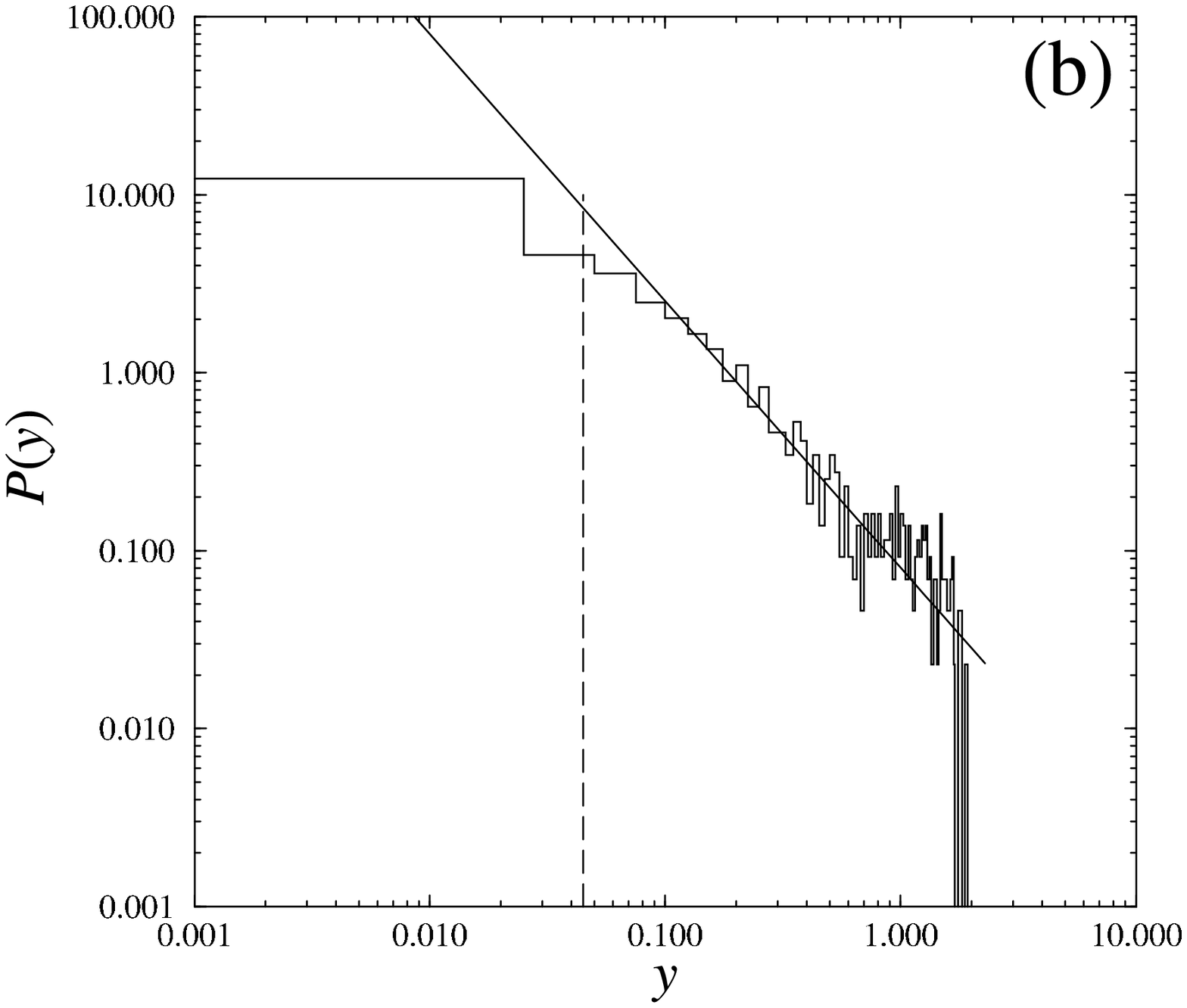}\newline
\includegraphics[width=6cm]{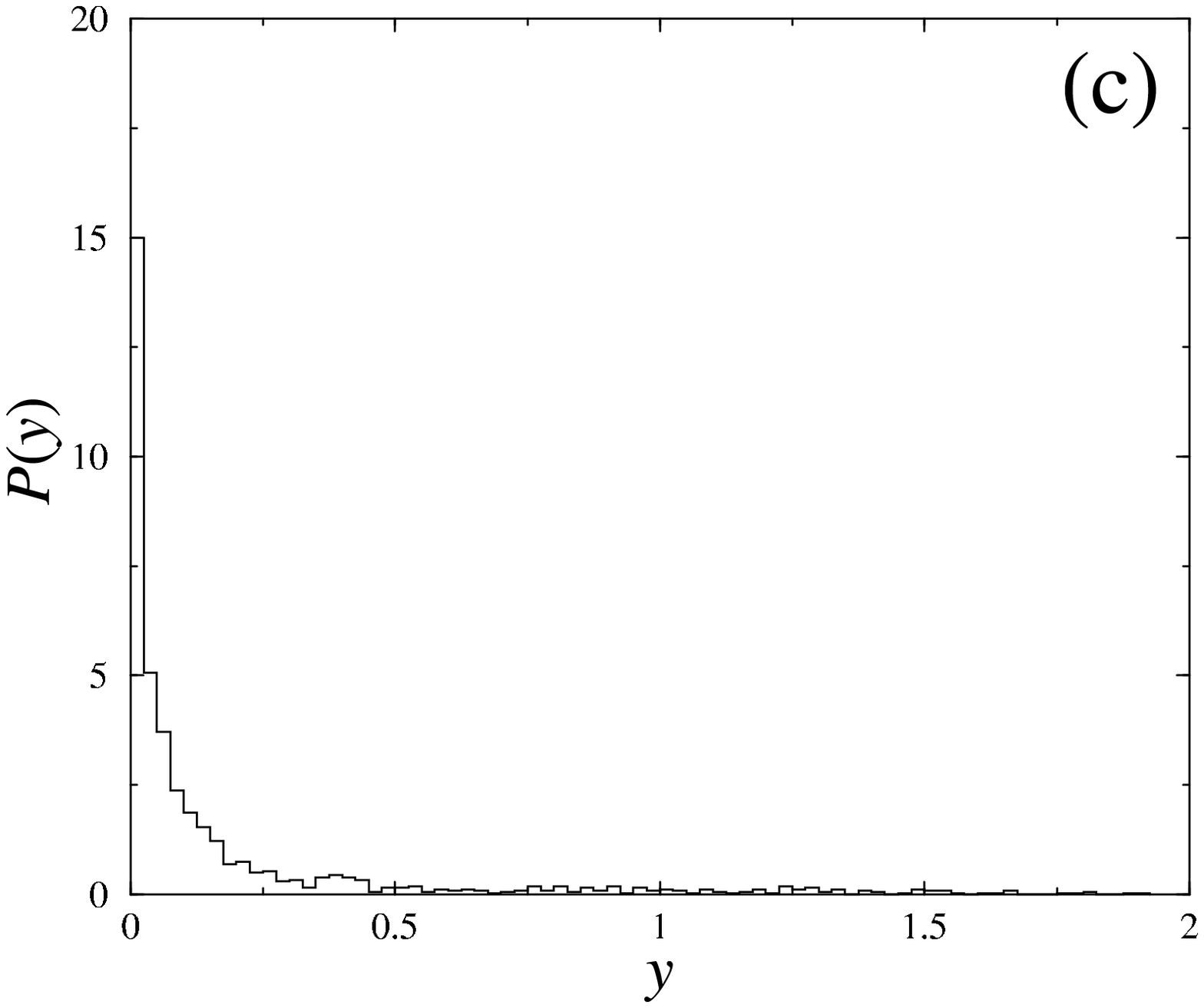}
\includegraphics[width=6cm]{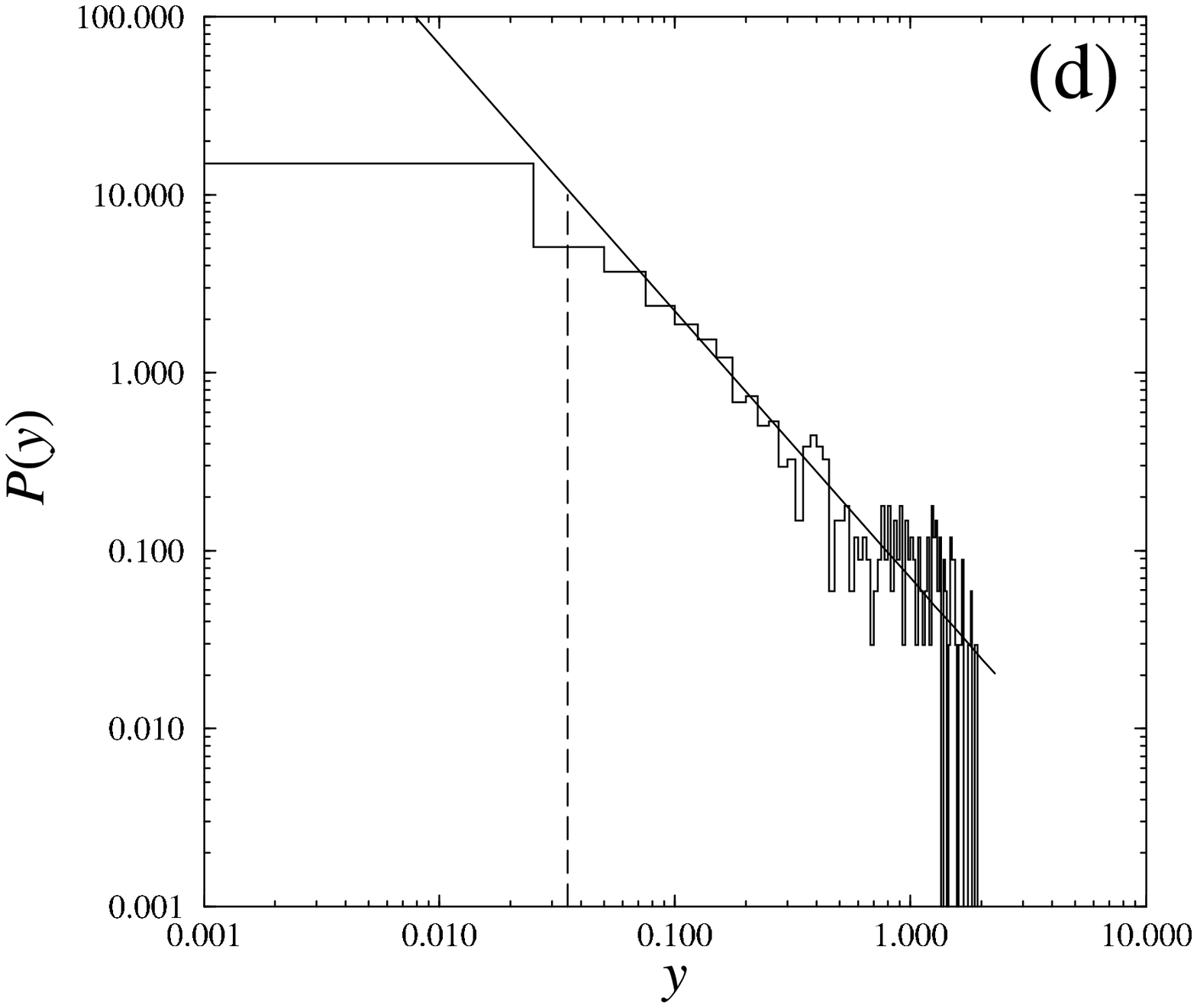}\newline
\caption{ $P(y)$ in normal and log-log scales for the chain of 
Fig. \ref{fig.chain}
with (a),(b) $N=6$; (c),(d) $N=7$. 
The continuous line shows the power law $y^{-3/2}$.
The distribution stops at the value $y_{\rm max}$. 
The dashed line indicates the value
$y=|\mathrm{Im}\,k|=\gamma_{\rm cl}/2$ associated with 
the classical escape rate $\gamma_{\rm cl}$. }
\label{pimk}
\end{figure}

For this graph, we have computed the density $P(y)$ of resonance widths
defined by Eq. (\ref{dfn.P(y)}).  The histogram of resonance widths is plotted
in Fig. \ref{pimk} for different sizes $N$ of the chain. Note that no resonance
appears below $y_{\rm max}=2.276$ which is the value computed from
Eq.(\ref{y_max}). This value is independent of the system size $N$ for $N\geq
3$. We have computed the eigenvalues of the matrix $\mathsf T$ for different
values of $N$ and we show the minimum eigenvalue in the following table:

\begin{center}
\begin{tabular}{|r|l|c|}
$N$ & $|\lambda _{\rm min}|$ & $y_{\rm max}$ \\ \hline
1 & 0.200 & 4.553 \\ 
2 & 0.392 & 2.655 \\ 
3 & 0.447 & 2.276
\end{tabular}
\end{center}

As we said for $N\ge 3$, this eigenvalue is independent of $N$ and is given
by the case with $N=3$.

Figure \ref{pimk} depicts the density of the imaginary parts
in a log-log scale and the power law $P(y)\sim y^{-3/2}$ conjectured in
Ref. \cite{Guarnieri} is observed. This is shown for chains of different
lengths in these figures. The power law holds for not to small sizes $N$. 
This observation is therefore a support for the conjecture of Ref.
\cite{Guarnieri} in the case of multiconnected graphs as the one of Fig.
\ref{fig.chain}.

\subsubsection{Detailed structures in the resonance spectrum}

Let us now discuss the details of the distribution of resonances.

\begin{figure}[th]
\centering
\includegraphics[width=8cm]{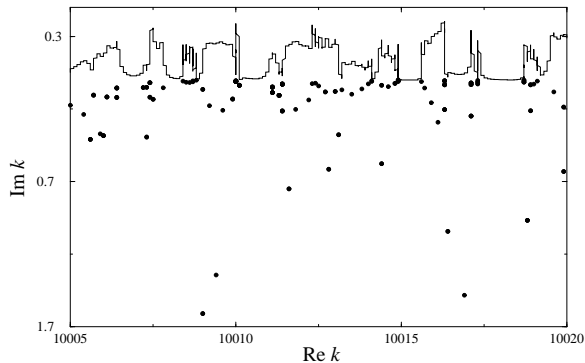}
\caption{Quantum scattering resonances (dots) and transmission probability
amplitude $T_{s_{8}s_{1}}$ (solid line) for the chain of Fig. 
\ref{fig.chain} with
$N=7$ unit cells.}
\label{fig.transres}
\end{figure}

First of all, we notice that the statistical description 
we have considered, that is the
distribution of $y$, is based on the homogeneous character of the resonance
distribution along the $x=\mathrm{Re}\,k$ axis. This homogeneity indeed holds
on large scales as we see in Fig. \ref{fig.res12}. Nevertheless, at small
scales, we can see in Fig. \ref{fig.transres} the formation of bands of
quantum resonances characteristic of a periodic system for which Bloch
theorem applies \cite{art1}.   The reason is that, for long times, the system
is able to resolve finer scales in wavenumber (or energy) and thus after a
given time the system will ``feel'' the periodic structure and
evolves ballistically as follows from Bloch theorem.

The distribution $P(y)$ shows a peak at the origin (i.e. $y\simeq 0$) which
grows with the number of unit cells, $N$. This behavior tells us that
``bands'' of resonances with small $y$ are created when we 
increase the size of the
chain. In fact we have that, as for the resonances of 
one-dimensional open periodic potentials \cite{art1}, 
we have $N-1$ resonances per band and that the bands
converge as $\sim 1/N$ to the real axis when $N$ increases. This is
consistent with the ballistic behavior that should be observed in
the long-time limit.

On the other hand, the resonances with larger values of $y=|\mathrm{Im}\,k|$
are not arranged in a band structure and their number does not increase when
we change the size $N$ of the system. In fact, these resonances are located
at the same position in the complex plane for every value of $N$. Therefore
we interpret such short-lived resonances as metastable states that decay
without exploring the whole system.

\begin{figure}[tbp]
\includegraphics[width=6cm]{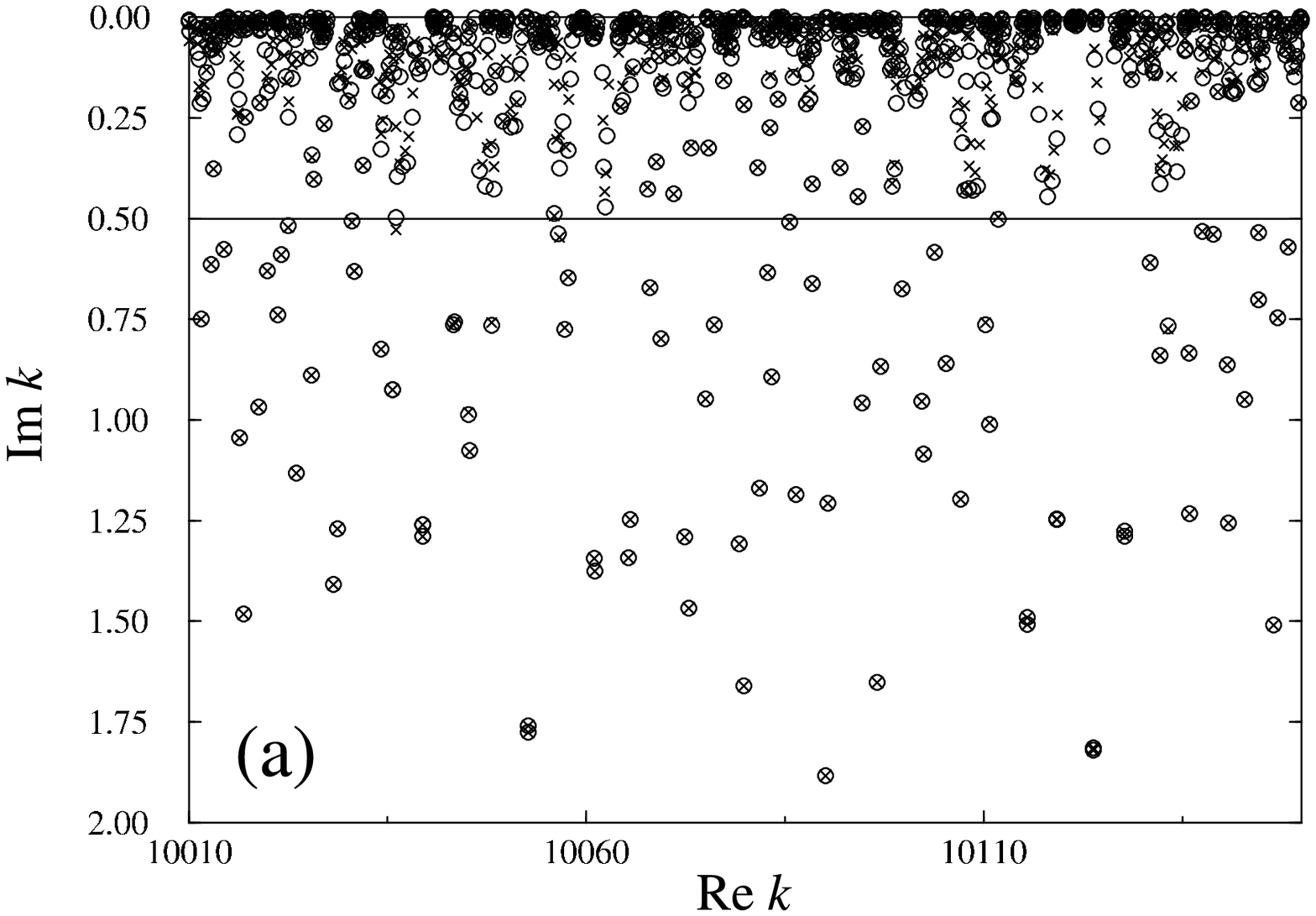}
\includegraphics[width=6cm]{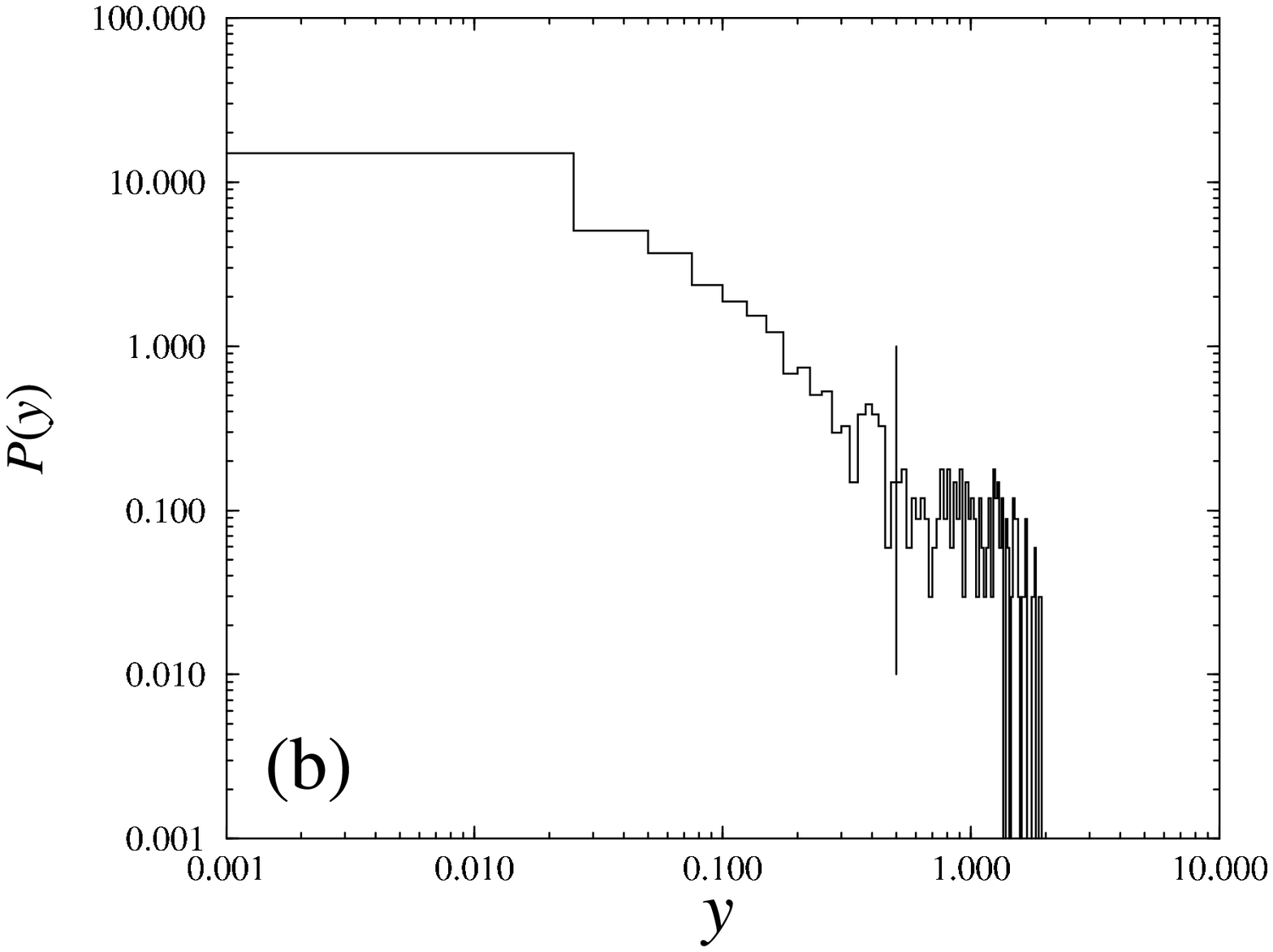}\newline
\caption{ (a) Resonances for the chain of Fig. \ref{fig.chain} with $N=7$ (circles)
and $N=8$ (crosses). (b) The distribution of resonance widths for the chain
with $N=7$. The straight line in figure (a) and (b) separates the two families
of resonances: those which depend on $N$ from those which do not. This figure
shows that the change in the distribution of the imaginary parts of the
resonances is associated  with the different families of resonances.}
\label{fig.desvs}
\end{figure}

In Fig. \ref{fig.desvs}a, we superpose the resonance spectrum for
chains with two different sizes. The figure shows that, for resonances with
large values of $y=|\mathrm{Im}\,k|$, neither their position, nor their number
change with $N$, but resonances with small values of $|\mathrm{Im}\,k|$
converges to the real axis as we increase $N$. 
Therefore, the relative number of resonances in the tail of the distribution
decreases as compared to the number of resonances with small values of $y$
which increases with $N$. We
can conclude that the resonance spectrum converges to the real axis in
probability when we increase $N$. This behavior is in contrast with the
simple systems analyzed in Ref. \cite {art1}, for which we observed that
\textit{each} resonance converges to the real axis as $N$ increases.

\begin{figure}[th]
\centering
\includegraphics[width=8cm]{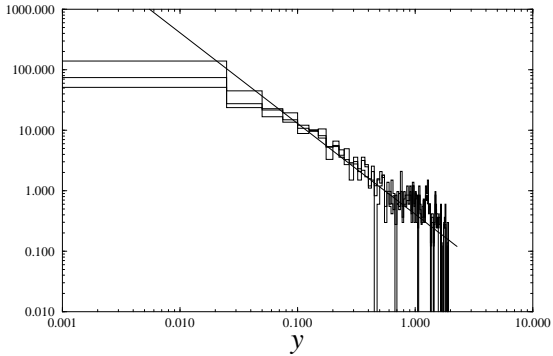}
\caption{Superposition of the log-log plots of 
$N\times P(|\mathrm{Im}\,k|)$ as a function of 
$y=|\mathrm{Im}\,k|$ for $N=5,6,8$ and the same graph as in Fig.
\ref{fig.chain}.}
\label{fig.escale}
\end{figure}

Eq.(\ref{res.dens.Guarnieri}) predicts a relative decrease of $P(y)$ as
$1/N$. This law is verified as shown in Fig. \ref{fig.escale} where we plot
for clarity the function $N\times P(\mathrm{Im}\,k)$ for only a few cases.
According to the theoretical distribution of Eq.(\ref{res.dens.Guarnieri}) 
$N\times P(\mathrm{Im}\,k) \propto\sqrt{D}|\mathrm{Im}\,k|^{-3/2}$, 
where the right-hand side is independent of $N$ and moreover, the
proportionality factor is determined by the diffusion coefficient of the
chain. We have computed the diffusion coefficient for this graph in Ref.
\cite{art4}. The continuous straight line in Fig. \ref{fig.escale}
corresponds to the diffusion coefficient of the infinite chain. The good
agreement shows that the proportionality constant $a$ in Eq.
(\ref{res.dens.Guarnieri}) is of order one, as expected.

In Fig. \ref{pimk}, we observe deviations from the power
law at small and large values of $|\mathrm{Im}\,k|$.
The dashed lines in Fig. \ref{pimk} indicate the 
value corresponding to the classical escape rate. 
We see from the figure that the
distribution of imaginary parts is well described by the power law for
$y=|\mathrm{Im}\,k|>\gamma_{\rm cl}/2$ and the classical escape rate
$\gamma_{\rm cl}$  is near the transition to this power law.
On the other hand, the tail of the distribution also deviates from the power
law. The distribution follows the power law until a value which decreases
when $N$ increases. Beyond this value the distribution seems to fluctuate
around an almost constant value, and then drops rapidly to zero. 

The region where the distribution of imaginary parts fluctuates around some
value corresponds to the region where the resonances are independent (in
number and in position) of the value of $N$. This can be seen in Fig. \ref
{fig.desvs}. In Fig. \ref{fig.desvs}a, 
we depict the resonances for $N=7$ and $N=8$ and
we draw a line which separate resonances that belongs to bands and change
with $N$ from those that are independent of $N$. This line is given by
$y=|\mathrm{Im}\,k|=0.5$. In Fig. \ref{fig.desvs}b,  we depict the density
of resonances for $N=7$, the value $y=|\mathrm{Im}\,k|=0.5$ is
indicated by the vertical straight line. This line marks the separation
between both families of resonances and thus the limit of the power law 
$y^{-3/2}$. 

\subsection{Linear graph with emerging diffusion}
\label{linear.graph}

For extended periodic open graphs, the classical decay
given by the leading Pollicott-Ruelle resonance,
corresponds to the decay of a diffusion process \cite{art4}. 
Here, we analyze the decay of the quantum staying probability
for such a graph. 

Here, we consider a linear periodic graph with a unit cell composed by two
bonds $a$ and $b$ of incommensurate lengths $l_a$ and $l_b$, respectively. At
the vertex that join these two bonds we have a scattering matrix
$\sigma(\eta_2)$ and the vertices that join two unit cells have a scattering
matrix $\sigma(\eta_1)$. These scattering matrices are of the form
\begin{equation}
\sigma(\eta)=\left[
\begin{array}{cc}
i\sin \eta & \cos \eta \\
\cos \eta & i \sin \eta
\end{array} \right]
\label{sigma.matrix}
\end{equation}
Therefore, the transmission and reflection probabilities for the classical
dynamics are $T_i=\cos^2(\eta _i)$ and $R_i=\sin^2(\eta_i)$.

\begin{figure}[th]
\centering
\includegraphics[width=12cm]{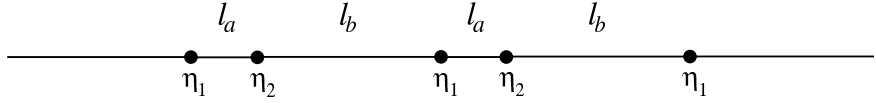}
\caption{Open graph built out of a periodic linear chain with a unit cell made
of two bonds with scattering matrices (\ref{sigma.matrix}) at each vertex. The
figure shows a chain with $N=2$ unit cells.}
\label{fig.lin.chain}
\end{figure}

Fig. \ref{fig.lin.chain} depicts an open graph by considering only $N$ units
cells connected to semi-infinite leads at the left-hand and right-hand side of
the finite graph.

\subsubsection{Classical diffusive behavior}

In the limit $N\to\infty$ of an infinitely extended chain, the 
graph becomes periodic and the motion is diffusive.  The  diffusion
coefficient can be calculated by considering the Frobenius-Perron-type matrix
${\mathsf Q}(s,q)$ defined in Ref. \cite{art4} in terms of the classical
wavenumber $q$ and the rate $s$.  For the present graph, we have that
\begin{equation}
{\mathsf Q}(s,q)=\left[
\begin{array}{cccc}
0 & e^{+iq}e^{-s\frac{l_b}{v}}T_1 &
e^{-s\frac{l_a}{v}}R_1 & 0\\ e^{-s\frac{l_a}{v}}T_2 & 0 & 0 &
e^{-s\frac{l_b}{v}}R_2 \\
 e^{-s\frac{l_a}{v}}R_2 & 0 & 0 & e^{-s\frac{l_b}{v}}T_2 \\
0 & e^{-s\frac{l_b}{v}}R_1 & e^{-iq}e^{-s\frac{l_a}{v}}T_1 & 0
\end{array}\right] \label{Qejem}
\end{equation}
where the columns and rows are arranged in the following order 
$(a,b,\hat{a},\hat{b})$. The diffusion coefficient is obtained 
from the second derivative of the first branch at $q=0$.
Developing $\det[{\mathsf I}-{\mathsf Q}(s,q)]$ for small values of $q$ and $s$
we get
$$
\det[{\mathsf I}-{\mathsf Q}(s,q)]=2l(R_1T_2+T_1R_2)\frac{s}{v}+T_1T_2q^2+
{\cal O}(s^2)+{\cal O}(sq^2)+{\cal O}(q^4)
$$
The diffusion coefficient $D$ is thus given by
\begin{equation}
D= -\left. \frac{1}{2}\frac{\partial ^{2}s_{0}(q)}{\partial q^2}\right| _{{\bf
q}=0}= \frac{vT_1T_2}{2l(R_1T_2+T_1R_2)} 
\label{Dejem}
\end{equation}
where $l=l_a+l_b$ is the total length of the unit cell. 

\begin{figure}[ht]
\centering
\includegraphics[clip=true,width=8cm]{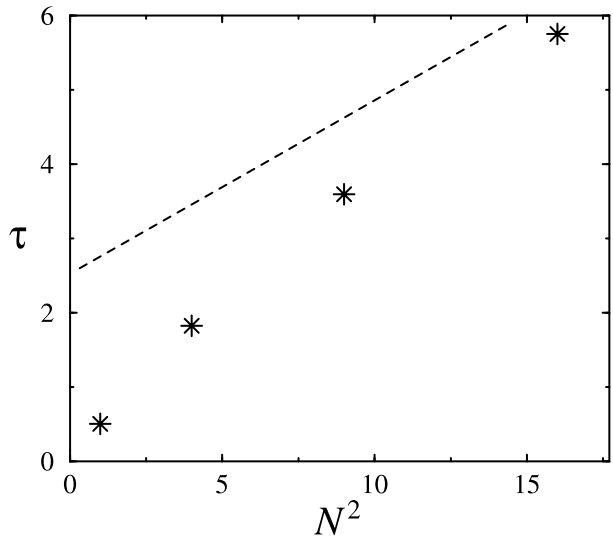}\\
\caption{Plot of the lifetime $\tau_{\rm cl}(N)=1/\gamma_{\rm cl}(N)$
for the chain of Fig. \ref{fig.lin.chain} with $N=1$, $2$, $3$ and $4$ unit cells as a function
of $N^2$. The dashed line corresponding to the diffusion coefficient
$D=0.506v$ given by Eq.(\ref{Dejem}) is approached by the classical 
lifetimes for increasing size $N$
of the open graphs.  We have used the parameters $\eta_1=0.1$ and
$\eta_2=(\sqrt{5}-1)/2$ and $l_a=0.5$ and $l_b=\sqrt{2}$.}
\label{Cdif.fig}
\end{figure}

In Ref. \cite{art4}, we have considered other examples and we have shown that
the classical escape rate $\gamma_{\rm cl}(N)$ for a finite open chain of size
$N$ is well approximated by Eq. (\ref{gamma}) in the limit $N\rightarrow
\infty$.  For the present linear graph, Fig. \ref{Cdif.fig} illustrates that
the classical lifetimes of the chain are indeed determined by the diffusion
coefficient according to $\tau_{\rm cl} =1/\gamma_{\rm cl} \sim N^2$ in the
limit $N\to\infty$.

\subsubsection{Spectrum of scattering resonances and its gap}

\begin{figure}[ht]
\centering
\includegraphics[clip=true,width=6cm]{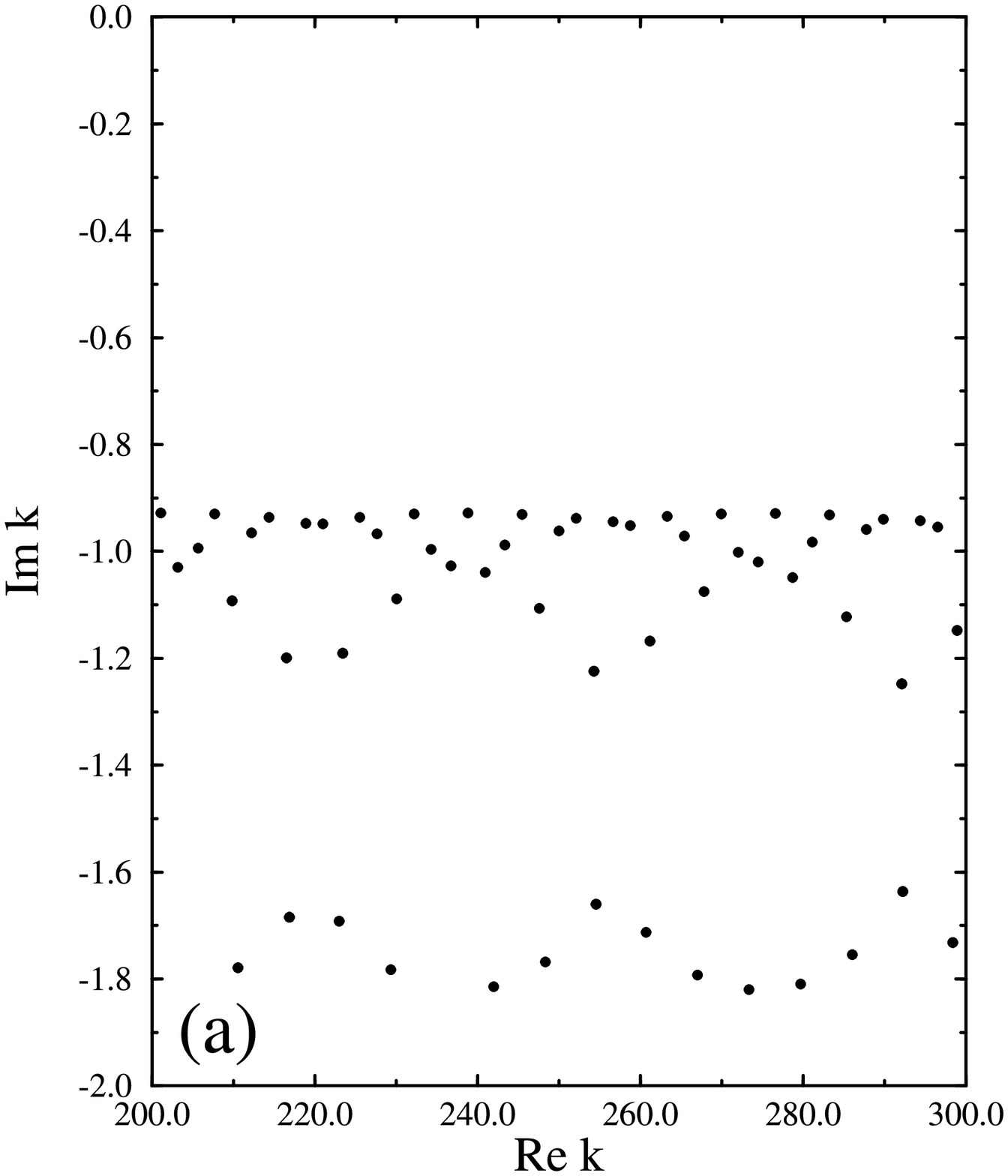}
\includegraphics[clip=true,width=6cm]{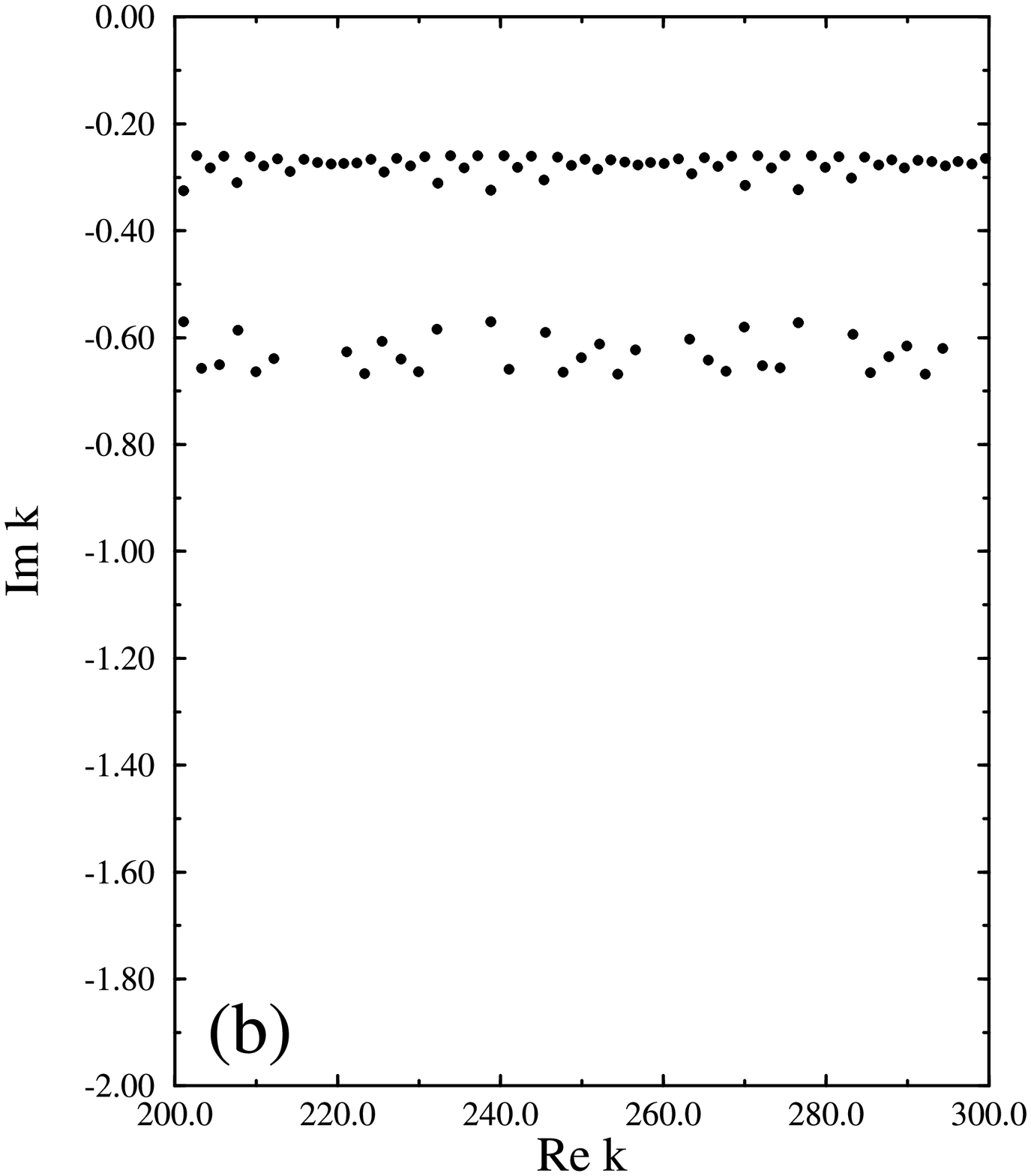}
\vskip 0.2cm
\includegraphics[clip=true,width=6cm]{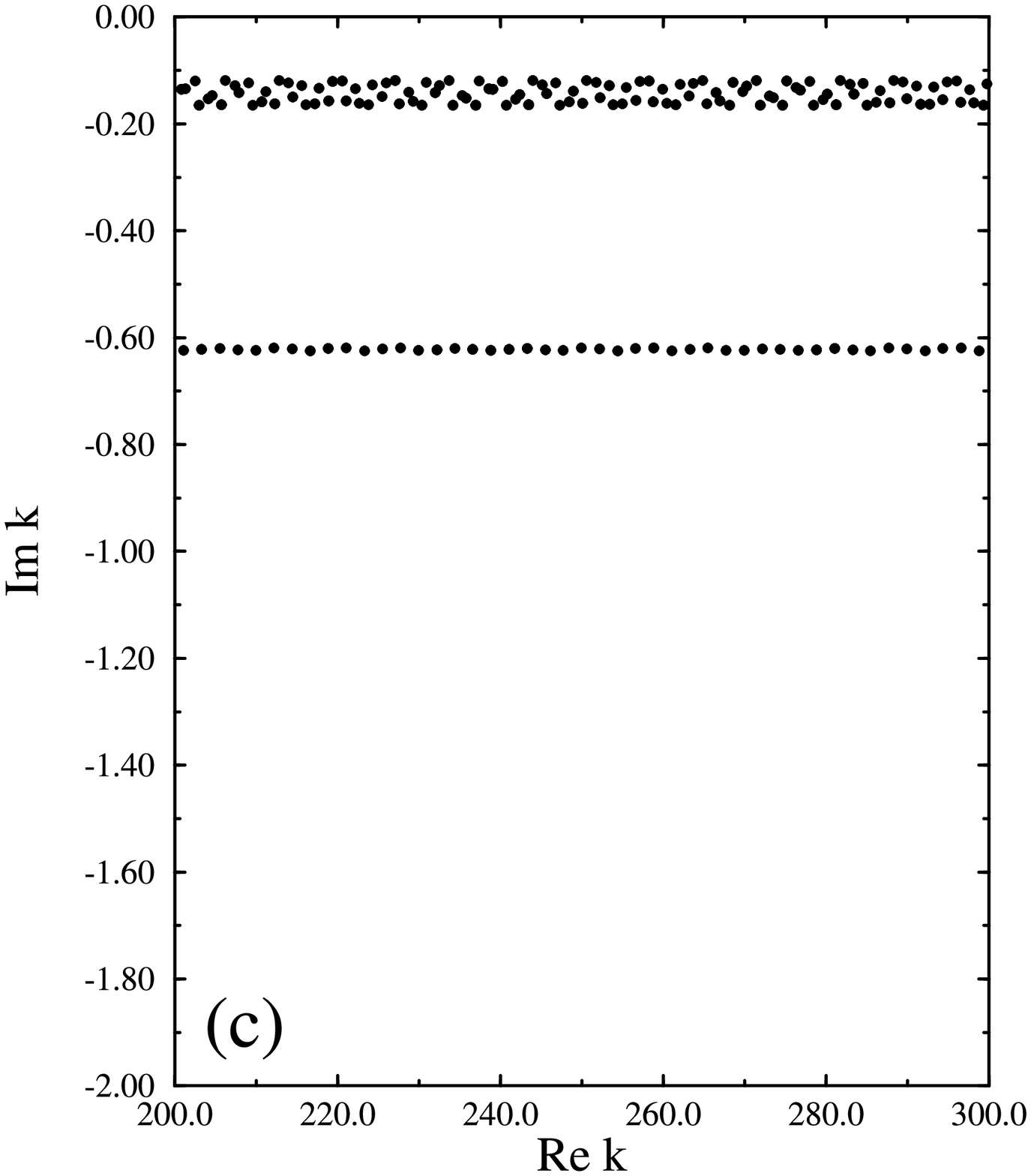}
\includegraphics[clip=true,width=6cm]{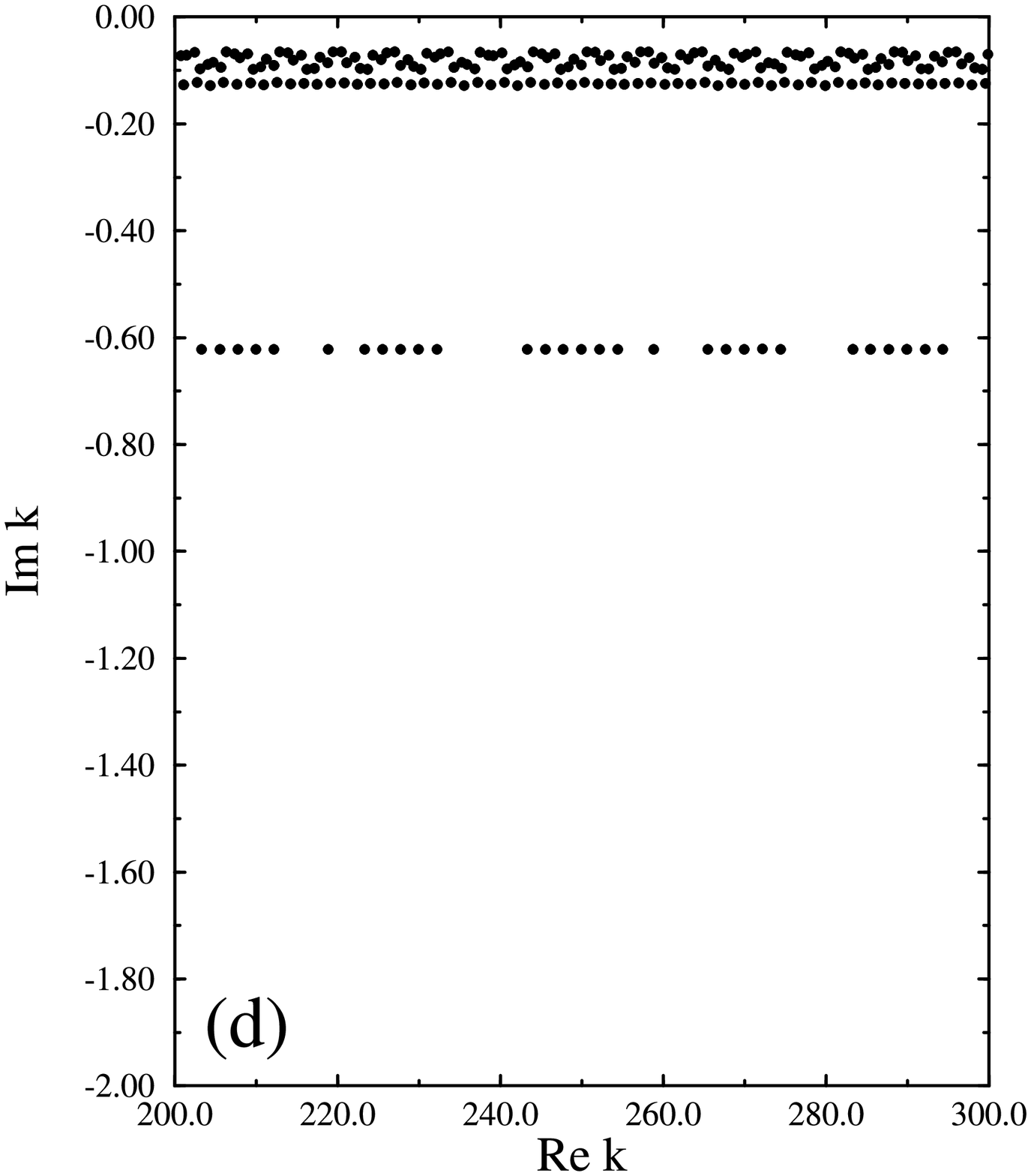}
\caption{Scattering resonances of the chain of Fig. \ref{fig.lin.chain} with
$N=1$, $2$, $3$ and $4$ unit cells.  We have used the parameters $\eta_1=0.1$
and $\eta_2=(\sqrt{5}-1)/2$ and $l_a=0.5$ and $l_b=\sqrt{2}$.}
\label{res9.fig}
\end{figure}

The resonance spectrum is depicted in Fig. \ref{res9.fig} for different chain
sizes $N=1$, $2$, $3$, and $4$.  

\begin{figure}[ht]
\centering
\includegraphics[clip=true,width=8cm]{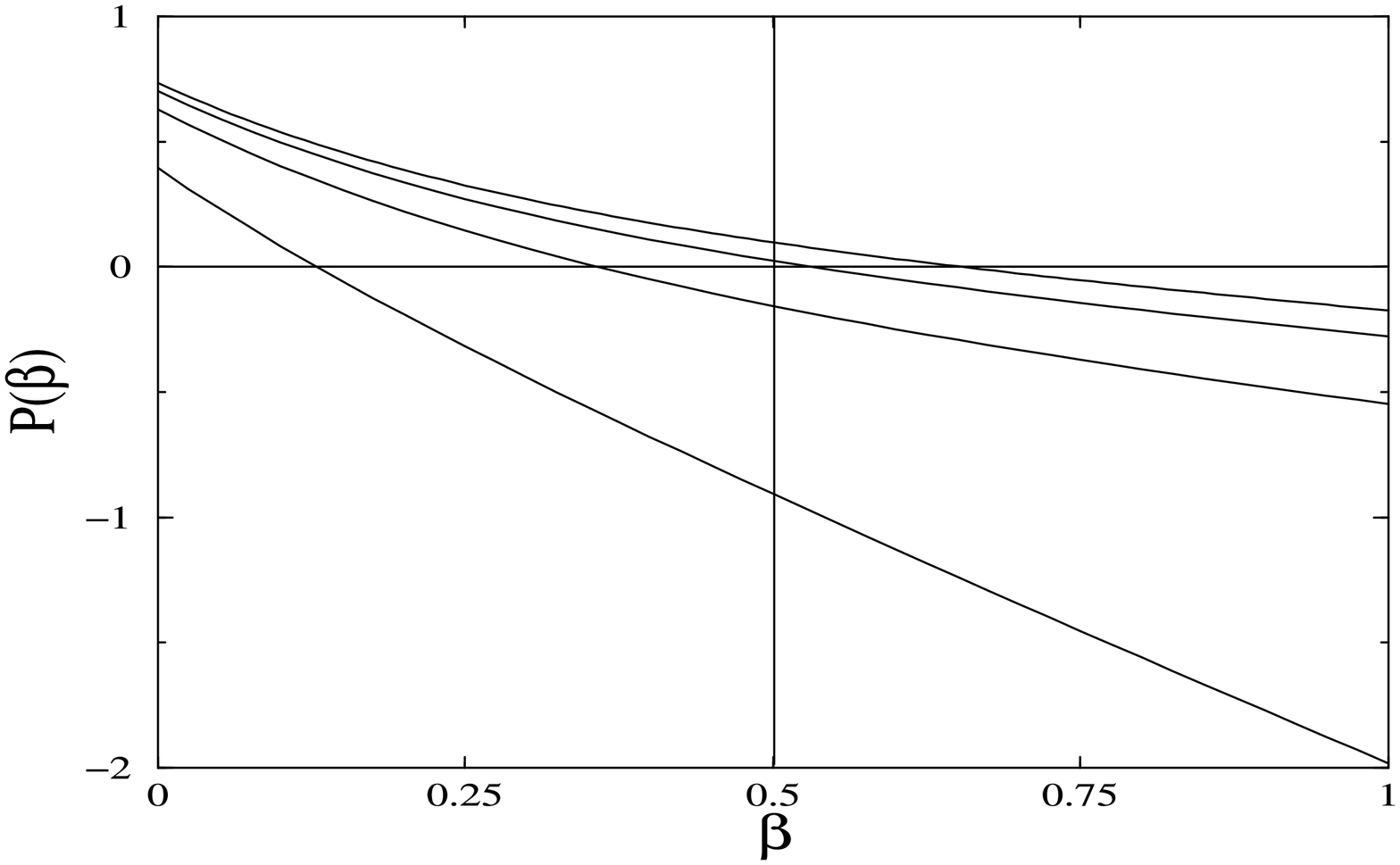}\\
\caption{From bottom to top, we plotted the topological 
pressure for the chain of Fig. \ref{fig.lin.chain} with the same parameter
values as in Fig. \ref{res9.fig} and the sizes $N=1$, $2$, $3$ and $4$ unit
cells, respectively. }
\label{presejem9.fig}
\end{figure}

The structures of the resonance spectrum and
the presence of a gap for the sizes $N=1$ and $N=2$ can be understood thanks to
the topological pressure plotted in Fig. \ref{presejem9.fig} for these graphs.
We see that the chains with $N=1$ and
$N=2$ unit cells has a value $\tilde P(1/2)<0$ and, therefore, a gap empty of
resonances. This gap appears in Figs. \ref{res9.fig}a and \ref{res9.fig}b.
For the chains with $N=3$ and $N=4$ unit cells, $\tilde P(1/2)>0$ 
and we do not have this upper bound for Figs. \ref{res9.fig}c and
\ref{res9.fig}d. 

\subsubsection{Decay of the quantum staying probability}

For these open graphs, we have calculated the quantum staying probability in
order to illustrate the emergence of a diffusive behavior.

The quantum time evolution has been computed by using the method described in
Subsection \ref{quant.evol.sec}.  The propagator was calculated by fast
Fourier transform from the Green function.  The staying probability was
calculated by considering an initial Gaussian wave packet located on a bond
$1$ inside the graph:
$$
\psi_{1}(y,0)=\left[\sqrt{\frac{2}{\pi}}
\frac{1}{\sigma (1-e^{-2\bar E\sigma^2})}\right]^{1/2} \sin(\sqrt{\bar E}y)
e^{-\frac{(y-y_0)^2}{4\sigma^2}}
$$
This wave packet is built by the modulation of a Gaussian with a sine function
of wavenumber $\sqrt{\bar E}$.  The wave packet is thus centered at $y=y_0$
(and we take $y_0=l_1/2$).  The wave packet is effectively on the bond 1 if 
$\Delta y =2\sigma << l_1$. 

\begin{figure}[ht]
\centering
\includegraphics[clip=true,width=8cm]{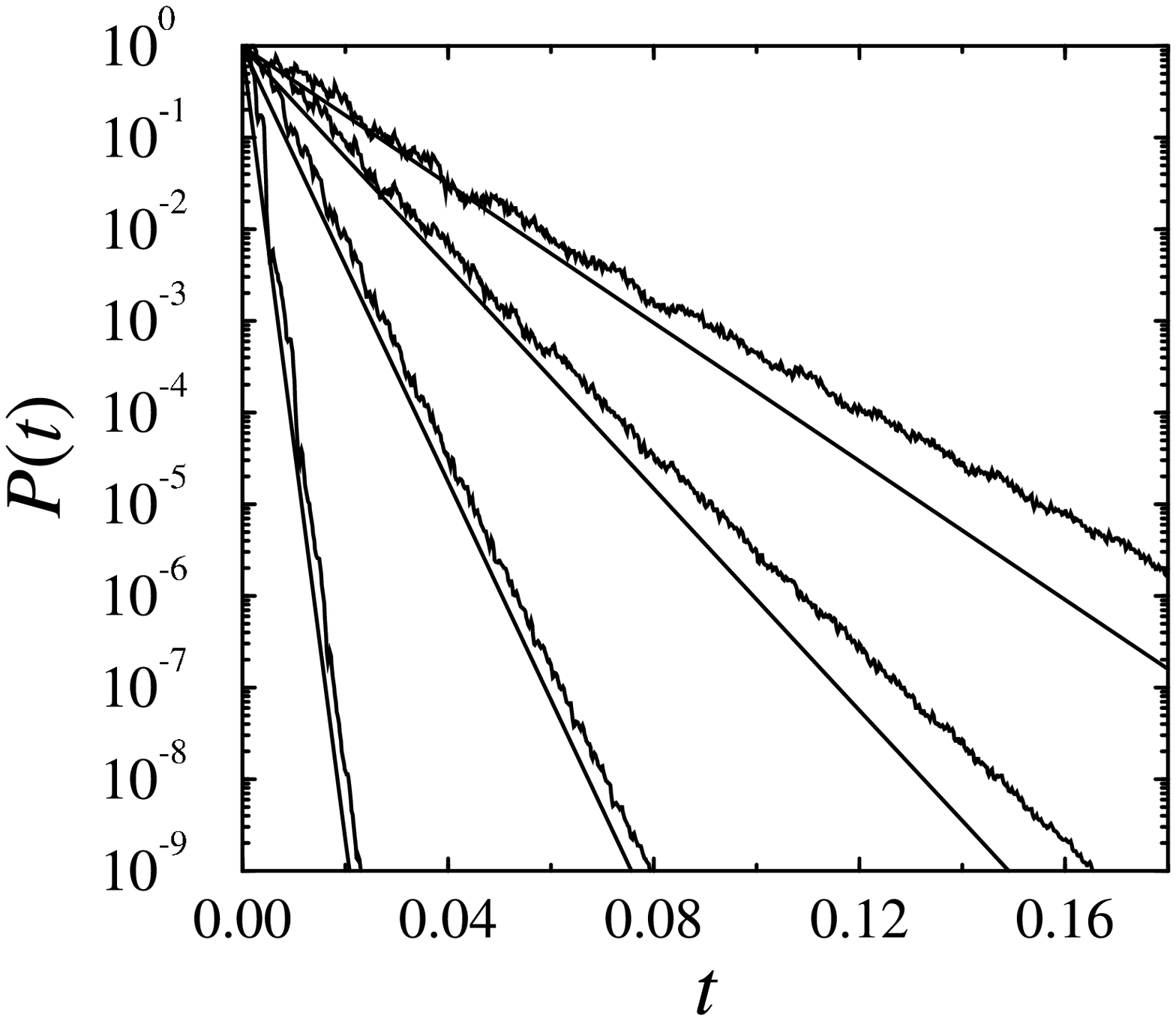}
\caption{From bottom to top: 
Decay of the quantum staying probability for the chain of Fig.
\ref{fig.lin.chain} with 1,2,3 and 4 unit cells and the same parameter values
as in Fig. \ref{res9.fig}. The straight lines gives the corresponding classical
decay as obtained from the Pollicott-Ruelle resonances. }
\label{decay2.fig}
\end{figure}

In Fig. \ref{decay2.fig}, we compare the decay of the quantum staying 
probability with the decay obtained from the leading Pollicott-Ruelle 
resonance for different sizes $N$ (see Fig. \ref{Cdif.fig}).  
This leading Pollicott-Ruelle resonance is
the classical escape rate $s_0=-\gamma_{\rm cl}(\bar v)$ at the velocity
corresponding to the mean energy $\bar E$ of the Gaussian wave packet.
The agreement observed in Fig. \ref{decay2.fig} 
for short times between the classical decay and
the quantum decay shows that for short times, 
the quantum evolution follows a diffusion process.

The deviation that appears in Fig. \ref{decay2.fig} 
for longer times corresponds again to the decay determined from an isolated
resonance and is therefore a pure quantum effect.

\section{Conclusion}
\label{conclu.sec}

We have studied dynamical and spectral properties
of open quantum graphs with emphasis in their relation
to the transport properties of the corresponding classical dynamics.
The classical dynamics has been obtained as the classical
limit of the quantum dynamics.

The time evolution of the quantum system is obtained
from the propagator which is computed as the Fourier transform
of the Green function. A closed expression and a multi-scattering
representation for the Green function has been presented.
We want to emphasize that both, classical and quantum evolution are
considered with the continuous time in opposition to the discrete time
evolution often considered in the literature of lattice networks.

In particular, we have computed the quantum staying probability.
For short times, this quantity decay exponentially in time with the 
classical escape rate. This classical escape rate
is obtained from the classical zeta function and correspond
to the leading Pollicott-Ruelle resonance. 
For large open periodic graphs the  leading Pollicott-Ruelle resonance
determines a decay dominated by diffusion and the decay of the quantum 
staying probability reveals at the quantum level the emerging diffusion
process. 

On the other hand, the quantum spectral properties are also related to
transport. The resonance spectrum  reveals some features related 
to both, the ballistic propagation
in the periodic system for the long time evolution and the diffusive 
classical dynamics that is an approximation for the quantum dynamics
for times shorter than the Heisenberg times. 

Indeed, the Fourier transform relates the long-time behavior
to the variations at small energy scales. 
At small scales, the resonance spectrum
is arranged in bands of $N-1$ resonances that converges as $\sim 1/N$
towards the real axis. Therefore the lifetime of each resonance in 
a band is proportional to the size of the system reflecting the 
ballistic transport that characterize periodic system as state the
Bloch theorem \cite{art1}.

Nevertheless, this ballistic propagation affects the long-time dynamics.
Indeed, the Bloch theorem gives information about stationary states.
The short time dynamics is determined by the large fluctuations of
the resonance spectrum. At a large scale, the resonance spectrum 
is uniform over the $\mathrm{Re}\,k$ axis (see Fig. \ref{fig.res12}).
This distribution display a power law $P(y)\sim y^{-3/2}$
early obtained in Ref. \cite{Guarnieri} and conjectured to be a general
law for open quantum systems with diffusive classical limit.
Our numerical results support this conjecture for multiconnected spatially
extended graphs.

Moreover, we have presented an alternative derivation of
the density of resonances obtained in Ref. \cite{Smilansky3}, which 
is based on general properties of the secular equations, namely, 
that is an almost-periodic function which in the appropriate
limit its mean motion is in the Lagrangian case. This  method
allows us to obtain an approximate lower bound for the distribution
of $y=|\mathrm{Im}\,k|$. Moroever, an upper bound for this distribution
has been obtained from the topological pressure $\tilde P(\beta)$.
This bound creates a gap in the resonance spectrum under the condition
$\tilde P(1/2)<0$ and is absent otherwise.

In summary, we have studied quantum properties of extended
open periodic graphs. We have shown that the Pollicott-Ruelle
resonances determine the decay of the quantum staying probability,
which in turn shows the appearance of diffusion at the quantum level.
The diffusion process is thus encoded in the distribution of 
scattering resonances. 

\section*{Acknowledgments}
The authors thank Professor G. Nicolis for support and
encouragement in this research.  Discussions with Professor M. Zworski are
acknowledged.  FB is financially supported by the ``Communaut\'e fran\c caise
de Belgique" and PG by the National Fund for Scientific Research (F.~N.~R.~S.
Belgium).  This research is supported, in part, by the Interuniversity
Attraction Pole program of the Belgian Federal Office of Scientific, Technical
and Cultural Affairs, and by the F.~N.~R.~S. .


\end{document}